\documentclass{article}
\pdfoutput=1 % if your are submitting a pdflatex (i.e. if you have
\usepackage[left=2.5cm, right=2.5cm, top=2cm, bottom=2cm]{geometry}
\usepackage{multirow}
\usepackage{multicol}
\usepackage{subcaption}
\usepackage{float}
\usepackage{placeins}

\usepackage{xcolor}

\usepackage{jcappub} % for details on the use of the package, please
\usepackage[T1]{fontenc} % if needed

\usepackage[nottoc]{tocbibind}

\title{A Critical Reanalysis of  Supernova Type Ia Data}
\author[a]{Ramanpreet Singh}
\author[a]{Athul C. N.}
\author[a,b]{H.K. Jassal}

\affiliation[a]{Indian Institute of Science Education and Research Mohali,\\
SAS Nagar, Mohali 140306, Punjab, India}
\affiliation[b]{National Centre for Radio Astrophysics, Tata Institute of Fundamental Research, Pune-411 007, India.}

\emailAdd{ph22043@iisermohali.ac.in}
\emailAdd{ms19027@iisermohali.ac.in}
\emailAdd{hkjassal@iisermohali.ac.in}

\date{}

\abstract{Cosmological parameter fitting remains crucial, especially with the abundance of available data. While many parameters have been tightly constrained, discrepancies—most notably the Hubble tension—persist between measurements obtained from different observational datasets.
In this paper, we re-examine the Pantheon supernova dataset 
to explore deviations in the distribution of distance modulus residuals from the Gaussian distribution, which is typically the underlying assumption. 
We do this analysis for the concordant cosmological constant model and for a variety of dynamical dark energy models.
It has been shown earlier that the residuals in this dataset are better fit to a logistic distribution. 
We compare the residual distributions assuming both Gaussian and Logistic likelihoods on the complete dataset, as well as various subsets of the data. 
The results, validated through various statistical tests, demonstrate that the Logistic likelihood provides a better fit for the full dataset and lower redshift bins, while higher redshift bins fit Gaussian and Logistic likelihoods similarly.
Furthermore, the findings indicate a preference for a cosmological constant model. 
However analysing individual surveys within the Pantheon dataset reveals inconsistencies among subsets. 
The level of agreement between surveys varies depending upon the underlying likelihood function.
}

\begin{document}
\maketitle

\section{Introduction}
\label{sec:intro}
A large number of observations have confirmed that the Universe is currently undergoing an accelerated expansion \cite{perlmutter1997measurements,Perlmutter_1999,Riess_1998} and this acceleration is being driven by a cosmological constant or by an exotic component called dark energy.
The cosmological constant and dark energy have the required negative pressure for the acceleration \cite{copeland2006dynamics}. 
While the cosmological constant model \cite{2024OJAp....7E..32B} is simplest and observationally the most favoured of these models, the problem of fine tuning which is associated with this model has motivated alternative formulations \cite{copeland2006dynamics,weinberg1989cosmological}.
If the dark energy equation of state deviates from that of cosmological constant model, the dark energy density varies with scale factor/redshift.
These models include those where dark energy equation of state parameter $w$ is $\neq -1$. 
The equation of state parameter may vary with redshift wherein one assumes a functional form of the parameter. 
Other descriptions of a varying dark energy equation of state parameter include  scalar field models, both canonical and noncanonical 
\cite{barreiro2000quintessence,ratra1988cosmological,kamenshchik2001alternative}. 
There have also been attempts to explain the accelerated universe via modified gravity models.
These include  $f(R)$ gravity theory \cite{de2010f,rippl1996kinematics, sotiriou2010f}, Modified Newtonian Dynamics (MOND) \cite{bekenstein2004relativistic,sanders1998cosmology},the dynamical dark energy model \cite{bagla2003cosmology,rajvanshi2019reconstruction,rajvanshi2021tachyonic,sangwan2018observational}.
The models are constrained well by observations though there is no compelling reason to favour one model over the other.

Cosmological parameters in the models are constrained by different and diverse observations.
The low redshift observations with which the cosmological models are compared with include observations of Supernovae Type Ia (SNe 1A), the observations which provided the first confirmation of the current accelerated expansion of the Cosmos 
\cite{Riess_1998,Perlmutter_1999}. The observational Hubble parameter \cite{2002ApJ...573...37J}, baryon acoustic oscillations \cite{2005ApJ...633..560E,adame2024desi6} , Local Dipole Anisotropy of deceleration parameter \cite{Rubin_2020} also support an accelerating Universe. Observations from the cosmic microwave background (CMB) radiation \cite{2016A&A...594A..13P} strongly aligns with these results and have provided tight constrains on the cosmological models.

While the cosmological parameters are being determined to better precision by current observations, more recently there has also come to light the so-called 'Hubble Tension". 
There is a discrepancy in the value of $ H_0 $ obtained from late time measurements (SNe IA, Cepheids, etc.) and early time measurements (CMB)\cite{riess2022comprehensive}
The value of Hubble constant  $ H_0 = 67.4 \pm 0.5 km s^{-1} Mpc^{-1} $ is obtained by fitting the $\Lambda$CDM model to the Cosmic Microwave Background (CMB) data \cite{aghanim2020planck,hinshaw2013nine,bennett2013nine}. Other surveys like Dark Energy Survey (DES) points towards the similar value of the Hubble constant\cite{camilleri2024dark} while putting some constraints on the other cosmological parameter as well \cite{ abbott2024dark,shlivko2024assessing}.
On contrary, a value $ H_0 = 73.04 \pm 1.04 km s^{-1} Mpc^{-1} $ \cite{2018ApJ...859..101S,scolnic2022pantheon+,riess2022comprehensive} is obtained from SH0ES and Pantheon/Pantheon+ analysis of Sne Ia and Cepheids, introducing a notable 4.4 to 6$\sigma$ discrepancy observed depending on the dataset considered. 

Several attempts are underway to explain this tension. 
The explanations for the tension and possible resolution attempts include introduction of early dark energy models \cite{niedermann2020resolving,poulin2019early}, modified gravity theories \cite{adi2021can,saridakis2023solving,10.1093/mnrasl/slad041}, Interaction between Dark Matter and Dark Energy \cite{montani2024kinetic}, etc.
Recent findings by Riess et al. \cite{riess2024jwst} show that combining JWST data yields $H_0$ estimates of \( 73.4 \pm 2.1 \), \( 72.2 \pm 2.2 \), and \( 72.1 \pm 2.2 \, \text{km/s/Mpc} \) for Cepheid, JAGB, and TRGB methods, respectively. The combined result of \( H_0 = 72.6 \pm 2.0 \, \text{km/s/Mpc} \) aligns closely with HST estimates, suggesting that systematic biases in the distance ladder are unlikely to resolve the Hubble Tension.
Whereas
Friedman et al. \cite{freedman2024status} present new results from the Chicago Carnegie Hubble Program (CCHP) using JWST data, yielding $H_0$ values of \( 69.85 \pm 1.75 (stat) \pm 1.54(sys) \, \text{km/s/Mpc} \) for Tip of the Red-Giant Branch (TRGB) stars, \( 67.96 \pm 1.85 (stat) \pm 1.90(sys) \, \text{km/s/Mpc} \) for J-Branch Asymptotic Giant Branch (JAGB) stars, and \( 72.05 \pm 1.86 (stat) \pm 3.10(sys) \, \text{km/s/Mpc} \) for Cepheids.
In \cite{gall2024hubble}, the authors propose an alternative approach to addressing the Hubble tension by focusing exclusively on "blue supernovae"—those with a colour parameter \( c < -0.1 \), which are less affected by dust extinction. By analysing only these supernovae in the Pantheon+ dataset, they derive a Hubble constant of \( 70.0 \pm 2.1 \, \text{km/s/Mpc} \), which is within \( 1\sigma \) of the Cosmic Microwave Background measurement. In contrast, using the full dataset yields a value of \( 73.5 \pm 1.1 \, \text{km/s/Mpc} \). This finding suggests that restricting the analysis to blue supernovae may help reconcile the Hubble tension.

Since there is no plausible theoretical explanation to the phenomenon, it is pertinent to explore other aspects of cosmological data analysis.
One way is to reevaluate the cosmological constraints using diverse  statistical tools at our disposal.
It was shown in \cite{DAINOTTI202430} that logistic and student's t-distribution is a better fit to residuals than the Gaussian distribution for Pantheon and Pantheon + data respectively. 
The focus is  on reducing the uncertainty of Hubble constant and matter density parameter by up to 40$\%$ by using Non-Gaussian distance moduli likelihood for parameter estimation.  
We revisit the analysis of supernova type Ia data to explore how the deviation from Gaussianity affects the allowed range of cosmological parameters and how binning of the data is relevant for this analysis.
The dataset we chose for this detailed analysis is the PANTHEON dataset, which consists of 1048 SNe Ia data with redshifts ranging from \(0.01 < z < 2.26\) \cite{2018ApJ...859..101S}. This dataset is a combination of various samples, including Center for Astrophysics CfA1-4 \cite{hicken2009cfa3, hicken2012cfa4}, the Carnegie Supernova Project \cite{ folatelli2009carnegie, stritzinger2011carnegie}, Panoramic Survey Telescope and Rapid Response System (Pan-STARRS1) \cite{chambers2016pan}, the Sloan Digital Sky Survey (SDSS) \cite{frieman2007sloan, kessler2009first}, the Supernova Legacy Survey (SNLS) \cite{conley2010supernova, Sullivan_2011}, the Supernova Cosmology Project survey (SCP) \cite{suzuki2012hubble}, Great Observatories Origins Deep Survey (GOODS) \cite{riess1998supernova}, and the Cosmic Assembly Near-infrared Deep Extragalactic Legacy Survey (CANDELS), Cluster Lensing And Supernova survey with Hubble survey (CLASH) \cite{graur2014type, rodney2014type}.

We take the approach that a more detailed and robust statistical analysis can potentially uncover underlying correlations and systematics. 
For this purpose, we revisit the SNe 1A data.
Typically, the Gaussian distance moduli likelihood is used to estimate cosmological parameters due to the central limit theorem. However, we also use a Logistic likelihood, which is a better fit for the Pantheon dataset as shown in \cite{DAINOTTI202430}. 
By revisiting the data with these assumptions, we show that the deviations from Gaussian behaviour are significant.
A Markov Chain Monte Carlo (MCMC) analysis reveals that the Logistic likelihood yields broader significance contours than the Gaussian likelihood with residuals as the random variable. 
These broader contours can possibly help reduce the Hubble tension, by way of  larger error bars.
We do this analysis with the full dataset and for different subsets of the data by way of binning it in redshift.
We also perform an independent analysis based on datasets from different surveys compiled in the Pantheon set, such as the SDSS \cite{frieman2007sloan, kessler2009first}, SNLS \cite{conley2010supernova, Sullivan_2011} and  Pan-STARRS1 (or PS1 in short) \cite{ 2014ApJ...795...45S}, to determine if there are systematic preferences for specific sets of parameters within different surveys.

This paper is structured as follows.  
After the introduction section \ref{sec:intro}, in the following section \ref{sec 2}, we discuss background cosmology and solutions with different parameterisations of dark energy equation of state parameter for a spatially flat universe. 
These different parameterisations includes models with constant or dynamical dark energy equation of state parameter. 
In section \ref{param fitting}, we have discussed the likelihood functions used and the statistical tools required to carry out the analysis.
In section \ref{sec 3}, the parameter estimation is done using method: 
% maximum likelihood analysis and
MCMC (Markov chain Monte Carlo) for Gaussian and Logistic likelihood and compare the results for the complete data set. In section \ref{sec 3.2}, we will bin the data in various non-overlapping redshift bins and do the same analysis as before. 
In section \ref{sec 4}, we compare constraints from  the \textit{viz}. SDSS, SNLS and PS1 datasets separately out of the full Pantheon compilation to check the consistency within these.

\section{Background Cosmology}
\label{sec 2}
A homogeneous and isotropic universe permeated with a  perfect fluid with  pressure $P$ and energy density $\rho$ is described by the Friedmann equations:
\begin{align}
   & \Big({\frac{\dot{a}}{a}}\Big)^2  =\frac{8\pi G}{3}\rho- \frac{k}{a^2}
   \label{F1}
    \\
    & \dot{\rho}+\frac{3\dot{a}}{a}(\rho+p) =0 \label{F2},
\end{align}
where the $a(t)$ is the scale factor and $k=\pm1,0$ determines the geometry of the Universe, a positive and negative value for closed and open universe respectively and is zero for a spatially flat universe. 
For the purpose of our analysis, we consider a spatially flat universe i.e $k=0$ which is well motivated by the observations of the Cosmic Microwave background Radiation \cite{aghanim2020planck,hinshaw2013nine,bennett2013nine}. 
We can model different components of the Universe as barytropic fluids, where the  pressure and density of the fluid are related as $p=w \rho$, where $w$ is  the  equation of state parameter. 
For the cosmological constant,  the equation of state parameter $w=-1$ \cite{peebles2003cosmological}. 
Cosmological constant is not the only possible 'perfect fluid' which can drive the accelerated expansion.
In general, the equation of state parameter can be a function of time. 
Simplest generalisation of $\Lambda CDM$ is $wCDM$ where the dark energy equation of state parameter is some constant $w=w_0 \neq -1$ \cite{tripathi2017dark,sangwan2018reconstructing}. 
Other generalisation is by way of assuming dark energy to be a perfect fluid with a dynamical equation of state parameter.

The lack of a theoretically motivated functions form of the fluid dark energy equation of state parameter, the dynamical nature of dark energy is modelled by way of parameterisations. 
Introduction of the Chevallier–Polarski–Linder (CPL)  parameterisation  $ w(z)=w_0+w^\prime {z}/{(1+z)} $ is well motivated by Linder \cite{chevallier2001accelerating,linder2003exploring} is bounded at low and high redshift. 
The Jassal-Bagla-Padmanabhan (JBP)  parameterisation $ w(z)=w_0+w^\prime {z}/{(1+z)^2} $  maintains a consistent equation of state both in the present epoch and at high redshifts, while undergoing rapid changes at low redshifts \cite{jassal2005wmap,jassal2005observational}. 
G. Efstathiou \cite{efstathiou1999constraining} introduced a new parameterisation of equation of state for dark energy $w(z)=w_0 + w'log(1+z)$ and it was demonstrated to fit well with observations in the redshift range  $z\leq 4$
and later modified by Lei Feng \textit{et. al.} \cite{feng2011new} to accommodate observations across a broader range of redshift. 
We introduce another parameterisation where the equation of state parameter $w(z)= (w_0 - w_1)(1+z)+w_1$ is a linear function of redshift. Such parameterisation has been previously considered by the SNAP collaboration \cite{aldering2002overview} given as $w(z)=w_0+w^\prime z$ which is just the  first order expansion of taylor series in $z$ about $z=0$. 
Linder \cite{linder2003exploring,linder2003probing} introduced the CPL parameterisation to extend dark energy parameterisation to $z>1$ since linear parameterisation grows unsuitably at $z>1$.
Also, it can be seen that all the other dynamical parameterisations exhibit linear variation at lower redshifts.

 The luminosity distance and the distance modulus \cite{copeland2006dynamics} to redshift $z$ is given by
\begin{equation}
d_L 
=\frac{c}{H_0} (1+z)\int^z_0 \frac{dz^\prime}{E(z^\prime)} 
\end{equation}
\begin{equation}
\mu_{th}=
5\log_{10}\left(\frac{d_L}{Mpc}\right)+25
\end{equation}
Type Ia supernovae (SNe Ia) alone cannot determine the Hubble constant \( H_0 \) due to their degeneracy with the absolute magnitude \( M \). However, \( H_0 \) can be estimated by fixing \( M \). Typically, Cepheid variable stars are used to determine \( M \) in case of SNe Ia. For instance, Scolnic et al. \cite{2018ApJ...859..101S} fixed \( M \) at -19.35 based on their equations, which resulted in \( H_0 = 70 \text{ km s}^{-1} \text{ Mpc}^{-1} \). 
In contrast, Riess et al.\cite{riess2022comprehensive} found \( M = -19.253 \pm 0.027 \) and \( H_0 = 73.04 \pm 1.04 \text{ km s}^{-1} \text{ Mpc}^{-1} \) by combining $42$ SNe Ia with Cepheid data from the same galaxies. The observed distance modulus $\mu_{obs}$ is directly taken from the Pantheon data (G10) itself which has been computed using the aforementioned values of M whose expression \cite{tripp1998two} is 
\begin{equation}
    \mu_{obs}=m_B -M +\alpha x_1-\beta c +\Delta_M+\Delta_B
\end{equation}
In this context, \( m_B \) denotes the observed B-band amplitude, \( x_1 \) is the stretch parameter, and \( c \) represents the colour. The parameters \( \alpha \) and \( \beta \) quantify the correlations between luminosity and \( x_1 \) and \( c \), respectively. The term \( M \) corresponds to the intrinsic B-band absolute magnitude for a supernova with \( x_1 = 0 \) and \( c = 0 \). Corrections to the distance estimate, namely \( \Delta_M \) and \( \Delta_B \), account for the effects of host galaxy mass and other systematic biases \cite{2018ApJ...859..101S}. 
Although there have been recent discussions about revisiting the standardisation of Type Ia supernovae, such as considering non-linearities in the stretch-luminosity relationship \cite{ginolin2024ztf,ginolin2024ztf1}, our analysis follows the methodology described by Scolnic et al. \cite{2018ApJ...859..101S}. For parameter estimation, it is necessary to specify the underlying cosmology which is mentioned in the following. 
From Friedman's equation (\ref{F1}), we have
\begin{equation}
H(z)=H_0 E(z^\prime)
\end{equation}
For different dark energy parameterisations considered in this paper, $E(z')$ is given as 
\begin{equation}
E(z^\prime)=  \begin{cases}
   \sqrt{\Omega_{m0}(1+z^\prime)^{3}+ 1- \Omega_{m0} } & \Lambda CDM
   \\
   \\
   \sqrt{\Omega_{m0}(1+z^\prime)^{3}+ (1- \Omega_{m0})(1+z^\prime)^{3(1+w_0)} } &  w CDM 
   \\
   \\
   \sqrt{\Omega_{m0}(1+z^\prime)^{3}+ (1- \Omega_{m0})(1+z^\prime)^{3(1+w_0+w^\prime)} \exp{\left(- \frac{3 w^\prime z^\prime}{1+z^\prime}\right)}}   
   &  CPL
   \\
   \\
   \sqrt{\Omega_{m0}(1+z^\prime)^{3}+ (1- \Omega_{m0})(1+z')^{3(1+w_0)} \exp\left( \frac{3w^\prime}{2} \left( \frac{z'}{1+z'}\right)^2 \right)} 
   &   JBP
   \\
   \\
   \sqrt{\Omega_{m0}(1+z^\prime)^{3}+ (1- \Omega_{m0})(1+z^\prime)^{3\left( 1+w_0 +\frac{w^\prime}{2} \log(1+z^\prime) \right)}} 
   &   log
    \\
    \\
    \sqrt{\Omega_{m0}(1+z^\prime)^{3}+ (1- \Omega_{m0})(1+z^\prime)^{3(1+w_1)} e^{ 3(w_0-w_1)z^\prime }  }.
    &   linear
\end{cases}
\end{equation}
For the purpose of our analysis,  we have neglected the contributions of radiation and curvature energy densities (the case of flat universe).
The cosmological parameters  are $H_0$, $\Omega_{m0}$, $w_0$, $w'$, and $w_1$. 
 Using several statistical tools 
 Markov Chain Monte Carlo (MCMC) method, we can constrain and estimate the allowed range of  these parameters.

\section{Cosmological Parameter Fitting}
\label{param fitting}
 The standard method of determining parameters is minimising $\chi^2$ \cite{sangwan2018reconstructing} which is defined as:
\begin{equation} 
\chi^2 = \sum_i \left(\frac{ \mu^{obs}_i- \mu^{th}_i}{\sigma_{\mu,i}}\right)^2
\end{equation}
The degrees of freedom (dof) of a data with number of data points (N) and number of parameters (m) is given by, $dof=N-M$.
$\chi^2_{dof}=\frac{\chi^2}{dof}\sim \mathcal{O}(1)$, indicates that the fit is acceptable, otherwise the errors are either underestimated or overestimated. 
 We use two probability distributions i.e.,  Logistic and Gaussian likelihood distributions:
  \begin{equation}
 \text{Logistic likelihood}
 = \prod_i\left( \frac{\pi}{4\sqrt{3}\sigma} sech^2\left( \frac{X_i-\langle X \rangle }{2\frac{\sqrt{3}}{\pi}\sigma}   \right)\right)
 \end{equation}
 and
  \begin{equation}
 \text{Gaussian likelihood}
 =\prod_i\left(\frac{1}{(2\pi)^{\frac{1}{2}} \sigma_X }   e^{-\frac{1}{2}\left( \frac{X_i-\langle X \rangle }{\sigma_X}   \right)^2} \right)
 \end{equation}
 where $X^i=\frac{\mu^i_{ob}-\mu^i_{th}}{\sigma^i _\mu}$. When we minimise the  $\chi^2$ for the Pantheon data,
 we get the values of the parameters say $\theta_{Gauss}$. Then treating $X=X_{\theta}$ as random variable, we have $mean(X_{\theta_{Gauss}})\approx 0.02$ and $\sigma(X_{\theta_{Gauss}})\approx 0.999$ for all parameterisations. Therefore we can take the approximation as $mean(X_{\theta_{Gauss}})\approx 0$ and $\sigma(X_{\theta_{Gauss}})\approx 1$ for all parameterisations, we get the Gaussian likelihood function (which is used in MCMC analysis) as

 \begin{equation}
 \text{Gaussian likelihood}
\approx \frac{1}{(2\pi)^{\frac{d}{2}}}   e^{-\chi^2/2}
 \end{equation}` 

Grid based searches for the minimum $\chi^2$ are appropriate when the number of parameters is small. 
If the number of parameters becomes large, more efficient methods of searching for a minimum $\chi^2$ need to be employed. 
One such method is the  Markov Chain Monte Carlo (MCMC) method.
MCMC algorithms draw samples $\theta_i$ in a random walk in the parameter space from the posterior probability density.
\begin{equation}
p(\theta|x) = \frac{1}{Z}(p(x|\theta)*p(\theta))
\end{equation}
where $p(\theta|x)$ is the posterior probability, $p(x|\theta)$ is the Likelihood Function, $p(\theta)$ is the prior probability, and $Z$ is the normalisation factor. 
Every point in the chain only depends on the position of the previous point.
The expectation value of the model parameter with $N$ samples is
\begin{equation}
\langle \theta \rangle = \int p(\theta|x) * \theta d\theta \approx \frac{1}{N}\sum_{i=1}^N \theta_i
\end{equation}

A crucial step in the analysis is to establish the convergence of a Markov Chain Monte Carlo (MCMC) analysis. 
The Gelman-Rubin convergence diagnostic \cite{gelman1992inference} provides a numerical summary that helps gauge the convergence status of multiple chains. 
If the chains have converged, then the mean of all chains together (inter-chain) and the mean of each chain (intra-chain) should agree within some tolerance.
The equations used for the GR convergence test, along with the results are provided in Appendix \ref{GRCT}.

We have utilised the \emph{emcee} package for Markov Chain Monte Carlo (MCMC) analysis in Python \cite{foreman2013emcee}. The package \emph{emcee} is designed to perform efficient sampling in multi-dimensional parameter spaces. The specific sampler employed in this analysis is the Ensemble Sampler, which handles sampling by running multiple parallel chains (walkers) that collectively explore the parameter space. This approach enhances convergence.
In our MCMC analysis, we have utilised $1,000$ walkers in the parameters space of each model and performed a total of $12,000$ steps, with the first $2,000$ steps allocated for burn-in. The remaining steps produced chains for each model parameter and we reported the mean and standard deviation of these parameters based on the results from these chains with Gaussian and Logistic likelihood.

We  use the Akaike Information Criterion (AIC), Bayesian Information Criterion (BIC) and Kolmogorov–Smirnov (KS) test for comparing the different likelihood models \cite{bozdogan1987model,akaike1983information}.
Mathematically, 
the BIC and AIC is given by
\begin{align}
    BIC & = -2log(L)+k\hspace{0.5mm}log(n)\\
      AIC & = -2log(L)+2k
\end{align} 
%helps to select the best model by balancing both goodness-of-fit and model complexity.
where $L$ is the likelihood function used, $k$ denotes the number of parameters in the model, and $n$ is the number of data points.
 Further,
using these AIC values obtained for different models, we can define a metric that calculates the relative likelihood of a model for the given dataset, called 'the Akaike weights' and is defined for $i^{th}$ model as \cite{wagenmakers2004aic}
\begin{equation}
    \delta_i = \frac{e^{-\frac{1}{2} \Delta_i}}{ \sum_j e^{-\frac{1}{2} \Delta_j}}
\end{equation}
where $\Delta_i=AIC_i -AIC_{min}$ and $\sum_i \delta_i=1$. 
The strength of evidence for one model over another is found by dividing their respective Akaike weights.
 The KS test \cite{an1933sulla,smirnov1948table}, a nonparametric test,  helps  determining if a sample matches a reference distribution and he p-value depends on the maximum difference observed between the cumulative distribution function of the sample and the reference distribution.

\section{SNe Type 1A Observations}
\label{sec 3}
SNe type 1A observations were the first ones to give a conclusive evidence of the acceleration of the Universe.
For our analysis, we revisit constraints on parameters of different models with Supernova data. 
We utilise $1048$ SNe Type Ia data in the "Pantheon sample"\cite{2018ApJ...859..101S}, which is a compilation of  data from various surveys. 
We take three broad approaches, one is to analyse the full data set and then different subsets of it based on different ways of binning.
Besides binning the entire sample by redshift, we can also divide it into five subsamples based on different surveys namely Pan-Starrs-1 (PS1), SDSS, SNLS, Low-z, and Hubble Space Telescope (HST). 
The low-z subsample includes a collection of smaller low-z surveys, while the HST subsample encompasses all HST surveys. 

\begin{figure}[ht]
\centering
\includegraphics[scale=0.28]{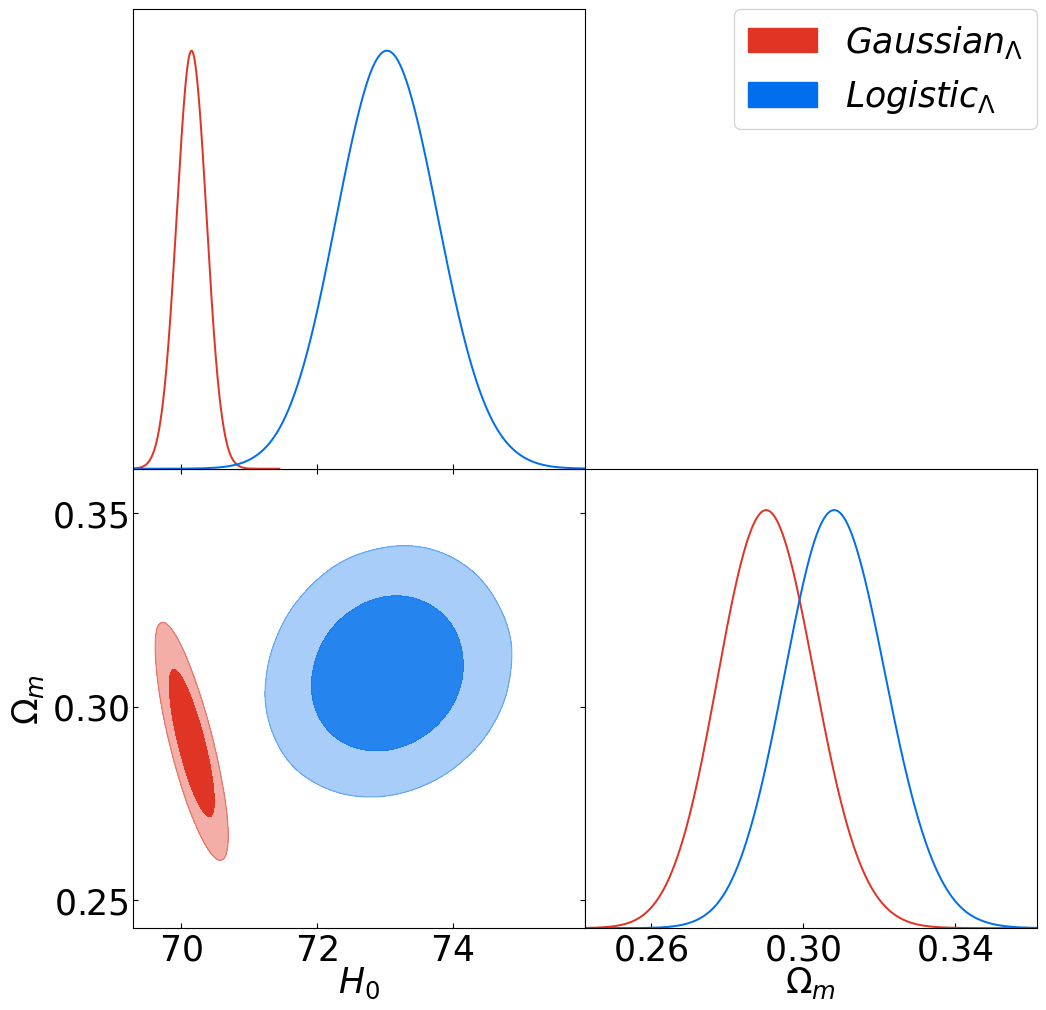}
\hfill
\includegraphics[scale=0.28]{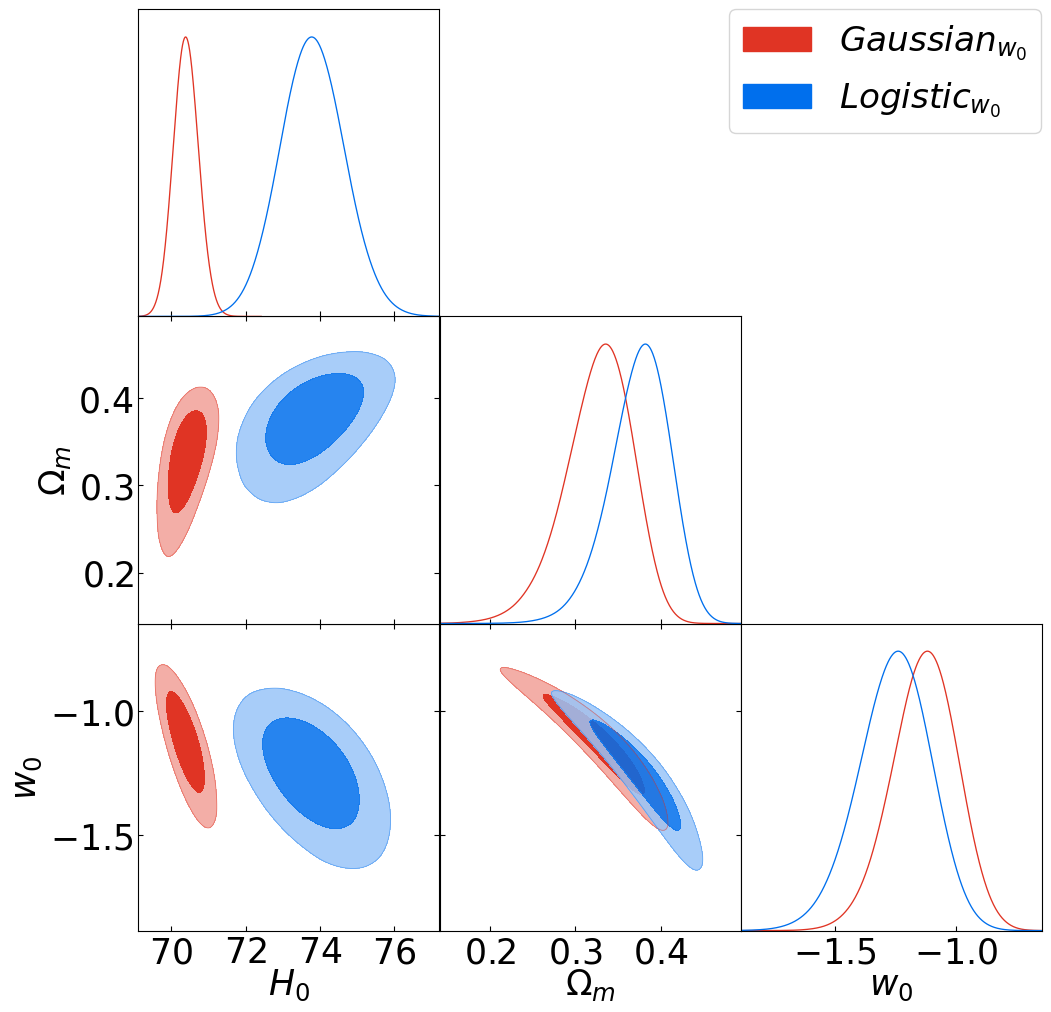}
\caption{Comparison of $1 \sigma$ and $2\sigma$ ranges for the cosmological constant model (left) and constant equation of state parameter $w$  model (right) using Gaussian and logistic likelihoods with the full dataset.  For these plots, it is evident that assuming a diagonal covariance matrix, the Gaussian likelihood is significantly more effective at constraining the parameters compared to the logistic likelihood. This trend holds irrespective of the parameterisation under consideration. The red contour represents Gaussian likelihood whereas the  blue contours show results for the logistic likelihood.}
\label{wcdmcon}
\end{figure}

We first analyse the Pantheon dataset as a whole as an independent analysis and to reconfirm results obtained earlier and to provide a reference for further analysis. 
Employing Gaussian and Logistic likelihood functions, our objective is to observe how the outcomes vary based on the selection of the likelihood.
The priors for the parameters are listed in table \ref{table:Param_bound}.
\begin{table}[t]
\centering
\begin{tabular}{|c|c|c|} 
    \hline
    Parameter &  Prior  \\ 
    \hline
    $H_0$ & $[50, 100]~km s^{-1} Mpc^{-1}$ \\ 
    \hline
     $\Omega_m $ & $[0.01, 1]$ \\ 
   \hline
    $w_0$   &$ [-2 ,-1/3]$ \\ 
    \hline
     $w^\prime$,$w_1$   &  $[-3,3]$ \\ 
    \hline
\end{tabular}
\caption{This table shows  the priors assumed for the parameters.}
\label{table:Param_bound}
\end{table}
The results of MCMC analysis are as given in Table \ref{mCmC for logistic} and the corresponding Gelman Rubin convergence test in Table \ref{GR for Logistic}.
We then  perform different tests: AIC, BIC and KS tests to check which likelihood is a better description of Pantheon data \ref{goodness od fit}. We found that the Logistic likelihood is preferred over Gaussian likelihood, showing agreement with \cite{DAINOTTI202430}. 
A comparison of the plots that are obtained from MCMC via considering Gaussian and Logistic likelihood in shown in fig. \ref{wcdmcon}.

\begin{table}[H]
\centering
\scalebox{0.80}{
    \begin{tabular}{|l|l|l|l|l|l|l|l|l|l|l|l|}
    \hline & Model    & $H_0$ & $\sigma_H$  &     $\Omega_{m0}$ &   $\sigma_{\Omega_m} $&     $w_0$ &$\sigma_{w0}$   &  $w^\prime $  & $\sigma_{w^\prime}$    & $w_1$  &$\sigma_{w_1}$  \\
    \cline{2-12}
\multirow{5}{5em}{Gaussian}&   $\Lambda CDM $ &70.157 & 0.220    & 0.291 & 0.013  & &  &   &    
 &       &  \\  \cline{2-12}
&    $w CDM $ &70.402 & 0.338   & 0.327 & 0.04  & -1.131& 0.135   &    
 &       & &   \\  
 \cline{2-12}
&    $CPL$ &  70.438 & 0.502   & 0.284 & 0.088  & -1.005 & 0.406  & 0.072 & 0.444 &    &  
\\ \cline{2-12}
&   $JBP$ &  70.353 & 0.419 &  0.350 & 0.111  & -1.129 & 0.139  &   -0.431  & 0.940  &     &   \\ 
\cline{2-12}
&   $log$ & 70.377  & 0.369  &  0.310  & 0.102   & -1.092   & 0.152  &  -0.356   &  1.136  &     &   \\ 
\cline{2-12}
&   $linear$ &  70.417 & 0.432 &    0.276  & 0.104 & -1.218  & 0.425  &    &     &  -1.369   & 0.907 \\
    
    \hline
     
  \multirow{5}{5em}{Logistic}   &      $\Lambda CDM$ & 73.04 & 0.74 & 0.31 & 0.01 & ~ & ~ & ~ & ~ & ~ & ~ \\ \cline{2-12}
      &     wCDM & 73.8 & 0.86 & 0.37 & 0.04 & -1.26 & 0.15 & ~ & ~ & ~ & ~ \\ \cline{2-12}
      &     CPL & 73.97 & 0.87 & 0.3 & 0.11 & -0.89 & 0.44 & 0.31 & 0.4 & ~ & ~ \\ \cline{2-12}
      &     JBP & 73.9 & 0.9 & 0.34 & 0.11 & -1.23 & 0.16 & 0.03 & 0.79 & ~ & ~ \\ \cline{2-12}
      &     Log & 73.87 & 0.87 & 0.32 & 0.11 & -1.17 & 0.19 & 0.23 & 0.94 & ~ & ~ \\ \cline{2-12}
      &     Linear & 73.85 & 0.87 & 0.32 & 0.12 & -1.29 & 0.7 & ~ & ~ & -1.43 & 1.46 \\ \hline
    \end{tabular}
    }
      \caption{Best fit value and $1\sigma$ deviation in parameters  obtained from MCMC analysis for the Gaussian and the Logistic likelihood}
      \label{mCmC for logistic}
\end{table}

%\FloatBarrier
\subsection{Data Subsets}
\label{sec 3.2}

\begin{figure}[ht]
\centering
\includegraphics[scale=0.5]{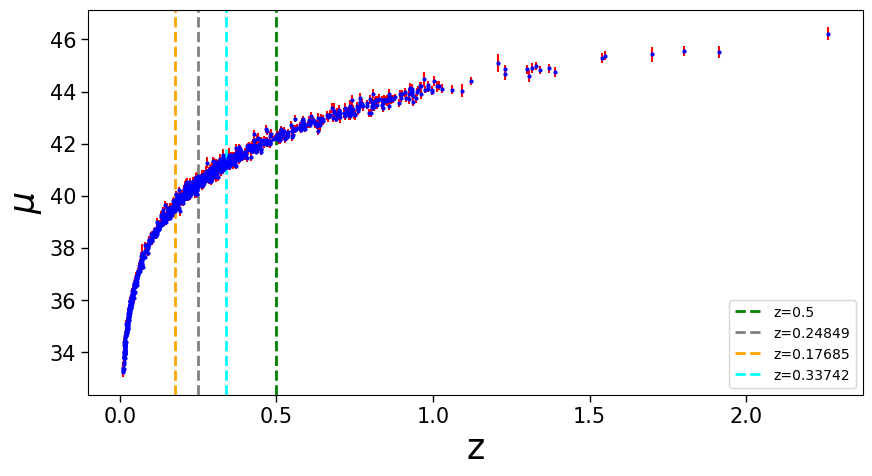}
\caption{This figure displays the boundaries of redshift bins employed for the Pantheon dataset. The $z\gtrless 0.5$ bins spans redshift range $z<0.5$ and $z>0.5$. For the case of two equal bins, the first bin spans the redshift range \(0.01012 < z < 0.24849\), while the second bin covers \(0.24862 < z < 2.26\). In the case of three equal bins,  the first bin covers \(0.0101 < z < 0.1769\), the second bin spans \(0.1771 < z < 0.3374\), and the third bin ranges from \(0.3375 < z < 2.26\). }
\label{binning}
\end{figure}

We divide the dataset into three different binning schemes. 
The first scheme separates the data into two bins: $z<0.5,~z>0.5$ where the redshift $z=0.5$ is chosen arbitrarily to check for trends in higher and lower redshifts.
The second scheme splits the data into two equal-sized bins, equal in terms of number of points. While the third scheme divides it into three equal-sized bins. These binning schemes are shown in fig. \ref{binning}.
 We then use MCMC method to estimate the optimised parameters for each individual bins and find allowed ranges of the parameters.

\begin{figure}[ht]
\centering
\includegraphics[scale=0.25]{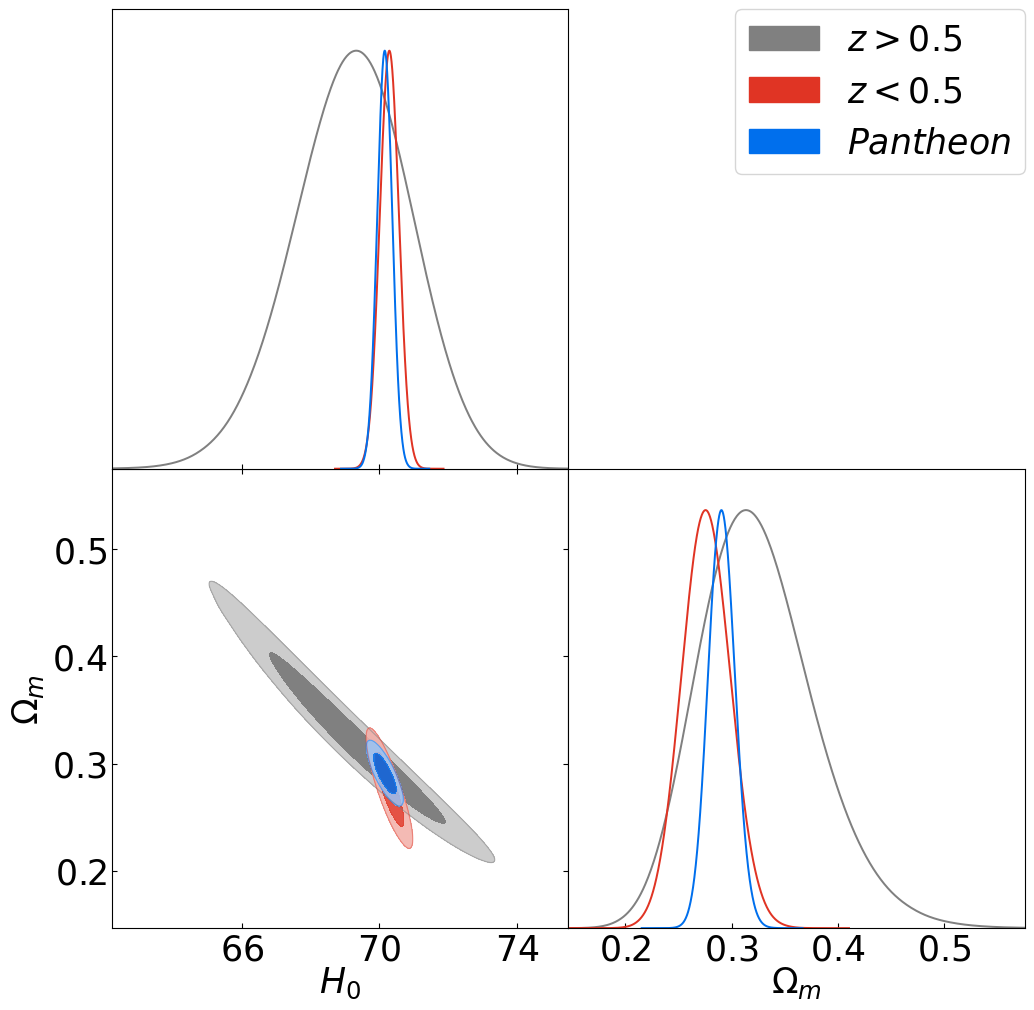} 
\hfill
\includegraphics[scale=0.25]{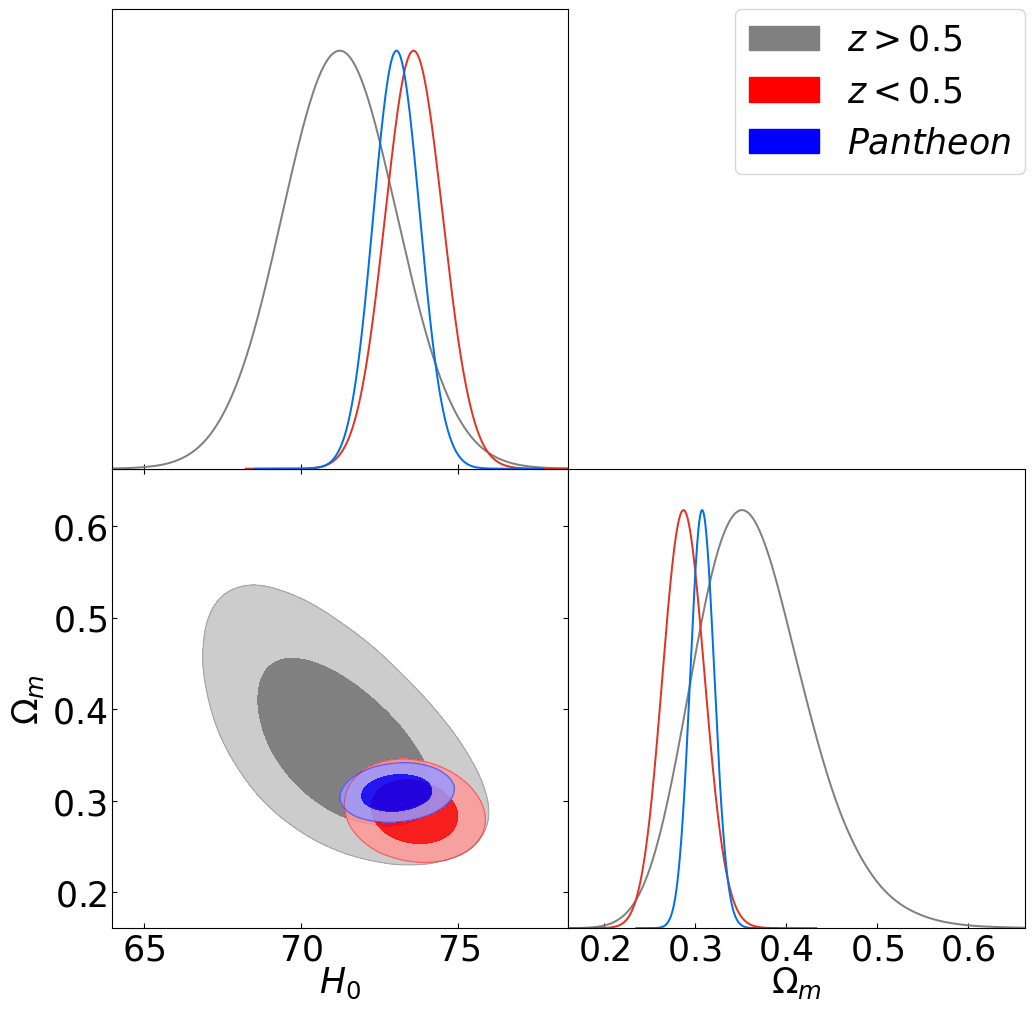}
\hfill
\includegraphics[scale=0.27]{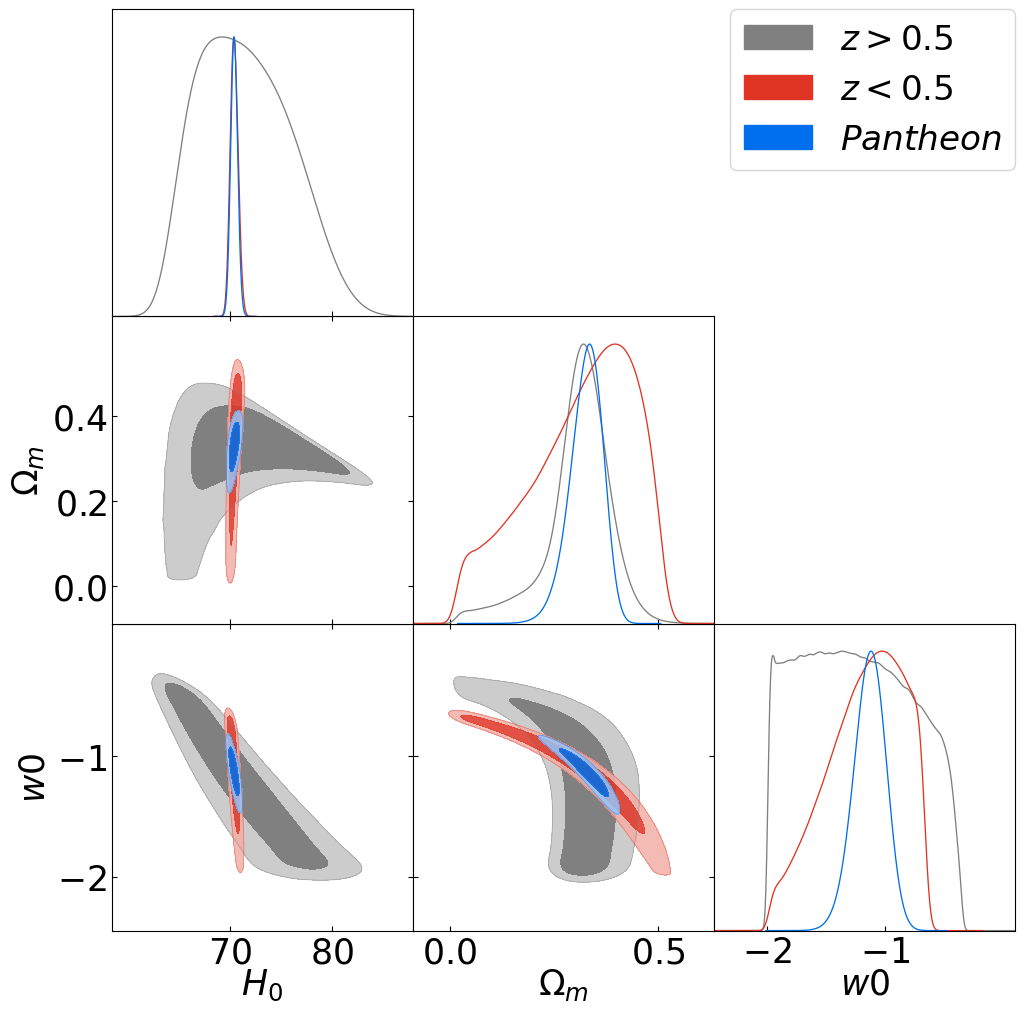} 
\label{w CDM_0.5 chi2}
\hfill
\includegraphics[scale=0.27]{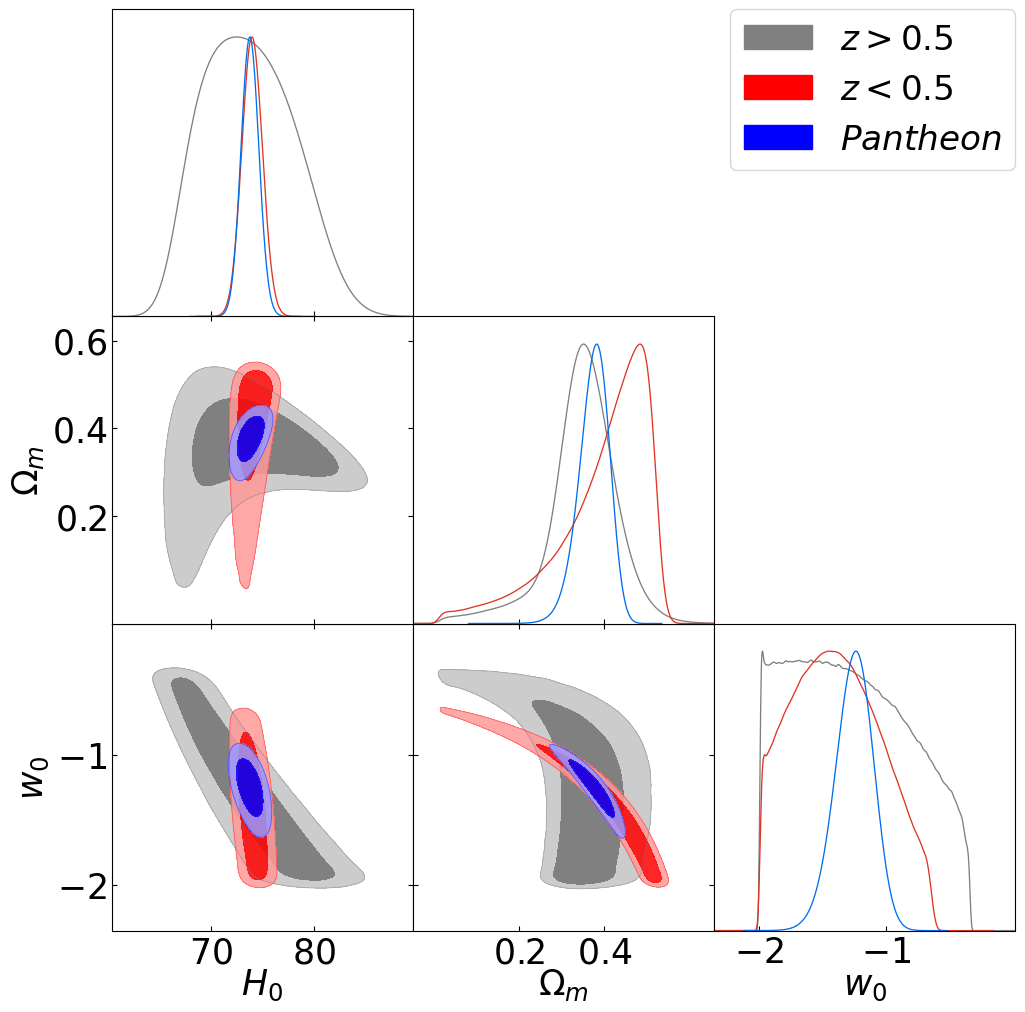}
\caption{The MCMC $2\sigma$ plots for the cosmological constant model (first row) and constant $w_0$ parameterisation (second row) with a Gaussian(first column) and Logistic (second column) likelihood using $z\gtrless 0.5$ Bins, indicate that the constraints are primarily influenced by the \( z < 0.5 \) supernovae, as the standard deviations for all parameters are smaller for this bin, likely due to it containing nearly four times as many data points as the \( z > 0.5 \) bin. The orientation of the contours can again be explained by examining the standard deviations of the parameters. It can be seen that the contours are broader in case of logistic likelihood compared to Gaussian likelihood. 
This trend is indeed followed by other parameterisation too as confirmed by the plots of constant $w_0$ parameterisation
Red contour represents $z<0.5$ bins; grey contour represents $z>0.5$ bins; blue contour represents the pantheon data. 
}
\label{wCDM_0.5-l}
\end{figure}

In the first binning scheme,
we choose $z=0.5$ as an arbitrary divide in low and high redshift with $832$,~$216$ data points in  $z<0.5$ and $z>0.5$ bin respectively. 
As demonstrated for the full dataset, the logistic likelihood provides a better description for the individual bins as well (the table of goodness of fit is \ref{goodness od fit}).
The results of MCMC analysis are as given in Table \ref{MCMC-l zmore 0.5}.
We then compare these bins with the full data set and verify the consistency of the full data set with the $z\gtrless0.5$ bins in $1$ (or $2$) $\sigma$ (see fig \ref{wCDM_0.5-l}) for both the likelihood functions.
It is clear that the constraints of the full data are being  predominantly due to the lower redshift data. The comparison shows that the  correlations between different parameters also follow the same trend as the data at lower redshifts. 
One  reason for this could be the higher number of data points in the lower redshift set. 

Splitting the Pantheon data in two equal halves according to the number of data points (524 data points in each bin), the first bin spans the redshift range $0.01012<z< 0.24849$, while the second bin spans the redshift in the range $0.24862 <z<2.26$. The results of MCMC analysis are as given in Table \ref{MCMC log 2nd half}. This again verifies the consistency of the full data set with  both the bins (see fig \ref{wCDM_0.5-l} for  the two  likelihood functions.
This binning serves the dual purpose of splitting the data in low and high redshifts and also to have enough data points in the two sections for  meaningful statistics. 
We can see that the higher redshift data set is comparatively more consistent with the whole data set. In terms of Hubble constant , both the bins are consistent for the cosmological constant model and the lower redshift prefers the  $w$CDM model.
For the matter density parameter, higher redshift data is consistent for both the models,  and for dark energy equation of state parameter ($w_0$), both the bins give similar results.

\begin{table}[t]
\centering
\scalebox{0.80}{
    \begin{tabular}{|l|l|l|l|l|l|l|l|l|l|l|l|l|}
    \hline  &  & Models  & $H_0$ & $\sigma_H$  &     $\Omega_{m0}$ &   $\sigma_{\Omega_m} $&     $w_0$ &$\sigma_{w0}$   &  $w^\prime $  & $\sigma_{w^\prime}$    & $w_1$  &$\sigma_{w_1}$  \\
   \cline{3-13}
\multirow{12}{5em}{Gaussian}  &  \multirow{6}{5em}{$z<0.5$ }&  $\Lambda CDM $ & 70.290  &  0.277   & 0.276 & 0.023  & &  &   &    
 &       &  \\ \cline{3-13} 
& &   $w CDM $ & 70.412   & 0.382   & 0.313 & 0.125  & -1.169 & 0.321   &  
 &       & &   \\  
\cline{3-13}
& &   $CPL$ & 70.509 &   0.592  & 0.335  & 0.131  & -1.081  & 0.451  &  0.178 & 0.549   &    &  
\\ \cline{3-13}
& &  $JBP$ &   70.443 &   0.446  & 0.369  & 0.154  & -1.267  & 0.319  &  -0.474 &  1.149  &    &  
\\  \cline{3-13} 
& &  $log$ & 70.485 &  0.419 & 0.337   &  0.130  & -1.235   &  0.312 &  -0.150 &  1.433 &    &  
\\  \cline{3-13} 
& &  $linear$  &  70.470 &  0.376   & 0.322 & 0.138  & -1.167  & 0.467  &   &    & -1.117   &  0.989
\\  \cline{2-13}

&  \multirow{6}{5em}{$z>0.5$ } &  $\Lambda CDM $ & 69.203 &  1.704   & 0.325  & 0.054  & &  &   &    
 &       &  \\ \cline{3-13} 
&  &   $w CDM $ &71.435 & 4.448   & 0.307 & 0.080  & -1.239 &  0.449 &    
 &       & &   \\  
 \cline{3-13}
&  &   $CPL$ & 75.561 & 10.377  &  0.287  &  0.083  &  -1.239  & 0.456 & 0.398  & 1.308 &    &  
\\\cline{3-13}
&  &  $JBP$ &  72.294 & 4.80  & 0.364    &  0.124  &  -1.246  & 0.466  & -0.796 & 1.384 &    &  
\\ \cline{3-13}
&  &  $log$ & 72.663  & 4.884  &  0.314  & 0.100   &  -1.282  & 0.447  & -0.382 & 1.594 &    &  
\\ \cline{3-13}
&  &  $linear$ & 79.673 &  9.561 &  0.293  & 0.083    & -1.167   & 0.479  &  &  &  -0.366  &  1.231
\\  \hline
  
	\multirow{12}{5em}{Logistic}&  \multirow{6}{5em}{$z<0.5$ }		&  $\Lambda CDM$ & 73.60 & 0.92 & 0.29 & 0.02 & ~ & ~ & ~ & ~ & ~ & ~ \\\cline{3-13}
			
		& 	&  wCDM & 74.01 & 0.99 & 0.40 & 0.11 & -1.39 & 0.34 & ~ & ~ & ~ & ~ \\\cline{3-13}
		& 	&  CPL & 74.36 & 1.05 & 0.42 & 0.12 & -1.09 & 0.46 & 0.56 & 0.57 & ~ & ~ \\ \cline{3-13}
			
			& &  JBP & 74.24 & 1.05 & 0.36 & 0.16 & -1.40 & 0.32 & 0.22 & 1.16 & ~ & ~ \\ \cline{3-13}
			
		& 	&  Log & 74.17 & 0.99 & 0.37 & 0.13 & -1.40 & 0.32 & 0.49 & 1.51 & ~ & ~ \\ \cline{3-13}
			
			& &  Linear & 74.13 & 1.03 & 0.41 & 0.14 & -1.24 & 0.47 & ~ & ~ & -0.98 & 1.09 \\ \cline{2-13}
	&  \multirow{6}{5em}{$z>0.5$ }	&  	$\Lambda CDM$ & 71.23 & 1.85 & 0.37 & 0.06 & ~ & ~ & ~ & ~ & ~ & ~ \\\cline{3-13}
	& 	&  	$wCDM$ & 73.67 & 4.31 & 0.35 & 0.08 & -1.27 & 0.45 & ~ & ~ & ~ & ~ \\ \cline{3-13}
	& 	&  	$CPL$ & 78.26 & 9.79 & 0.32 & 0.08 & -1.24 & 0.46 & 0.51 & 1.26 & ~ & ~ \\\cline{3-13}
	& 	&  	$JBP$ & 74.53 & 4.59 & 0.40 & 0.12 & -1.27 & 0.46 & -0.85 & 1.38 & ~ & ~ \\ \cline{3-13}
		& 	&  $Log$ & 74.98 & 4.74 & 0.34 & 0.11 & -1.3 & 0.44 & -0.28 & 1.66 & ~ & ~ \\ \cline{3-13}
		& 	&  $Linear$ & 81.39 & 8.91 & 0.32 & 0.09 & -1.17 & 0.48 & ~ & ~ & -0.37 & 1.25 \\ \hline
			
		\end{tabular}
  }
		\caption{This table shows the optimised parameter values obtained via MCMC analysis for $z \gtrless 0.5$. When compared with \ref{mCmC for logistic}, it is evident that the \( z < 0.5 \) bin produces results similar to those obtained from the full Pantheon dataset. }
		\label{MCMC-l zmore 0.5}
\end{table}

\begin{table}[ht]
\centering
\scalebox{0.80}{
    \begin{tabular}{|l|l|l|l|l|l|l|l|l|l|l|l|l|}
    \hline&  &   Model & $H_0$ & $\sigma_H$  &     $\Omega_{m0}$ &   $\sigma_{\Omega_m} $&     $w_0$ &$\sigma_{w0}$   &  $w^\prime $  & $\sigma_{w^\prime}$    & $w_1$  &$\sigma_{w_1}$  \\
    \cline{3-13}
 \multirow{13}{5em}{Gaussian} & \multirow{5}{5em}{$\textbf{I}st$-Half }  &  $\Lambda CDM $ & 70.316 &  0.353   & 0.266  & 0.046   & &  &   &    
 &       &  \\   \cline{3-13}
&  &   $w CDM $ &  70.444 & 0.389   & 0.398 &  0.140 &  -1.410 & 0.375 &    
 &       & &   \\  
 \cline{3-13}
&  &   $CPL$ &  70.839 &   0.506  &  0.567 &  0.136  & -1.187 & 0.486  & 1.449  & 0.967  &    &  
\\   \cline{3-13}
&  &  $JBP$ &  70.644   & 0.459  &  0.352  &  0.158  &  -1.469 & 0.327   &   0.888 &   1.537 &     &   \\ 
 \cline{3-13}
&  &  $log$ & 70.490    & 0.398  &  0.383   &  0.148  & -1.415  &  0.361  & 0.515   & 1.688   &     &   \\ 
 \cline{3-13}
&  &  $linear$ &  70.981  & 0.567 &  0.606   &  0.135 &  -1.258  &  0.466  &    &     & 0.648   &  1.416  \\   \cline{2-13}

&  \multirow{5}{5em}{$\textbf{II}nd$-Half } &  $\Lambda CDM $ & 70.04  &  0.550   & 0.296  & 0.022  & &  &   &    
 &       &  \\   \cline{3-13}
&  &   $w CDM $ &72.166 & 1.890   & 0.337 &  0.048  & -1.399 & 0.332   &    
 &       & &   \\  
 \cline{3-13}
&  &   $CPL$ &  73.252  &  3.554   &  0.324  & 0.084    &  -1.228   &   0.461 & 0.365  & 0.853 &    &  
\\   \cline{3-13}
&  &  $JBP$ & 71.945 &  2.661 & 0.367  &  0.124  &  -1.319 & 0.401  &   -0.582 &  1.465 &     &   \\ 
 \cline{3-13}
&  &  $log$ & 72.316 &  2.176  & 0.324    & 0.086  &  -1.364  & 0.355  &  -0.184  &  1.553   &     &   \\   \cline{3-13}
&  &  $linear$ & 73.119  & 3.249   &  0.325   & 0.088  &  -1.266  & 0.475  &    &     &   -0.929  &  1.084 \\   \cline{3-13}
   
			 \cline{1-13}

	 \multirow{13}{5em}{Logistic } & \multirow{5}{5em}{$\textbf{I}st$-Half }  & 	$\Lambda CDM$ & 73.36 & 1.23 & 0.26 & 0.04 & ~ & ~ & ~ & ~ & ~ & ~ \\  \cline{3-13}
		& & 	wCDM & 73.82 & 1.28 & 0.44 & 0.12 & -1.56 & 0.34 & ~ & ~ & ~ & ~ \\  \cline{3-13}
		& & 	CPL & 75.26 & 1.40 & 0.64 & 0.07 & -1.27 & 0.49 & 2.18 & 0.71 & ~ & ~ \\ \cline{3-13}
		& & 	JBP & 74.69 & 1.36 & 0.29 & 0.15 & -1.52 & 0.28 & 1.92 & 1.1 & ~ & ~ \\  \cline{3-13}
		& & 	Log & 73.97 & 1.29 & 0.40 & 0.14 & -1.52 & 0.34 & 0.98 & 1.61 & ~ & ~ \\  \cline{3-13}
		& & 	Linear & 76.49 & 1.59 & 0.69 & 0.05 & -1.52 & 0.38 & ~ & ~ & 1.93 & 0.93 \\  \cline{2-13}

		&\multirow{5}{5em}{$\textbf{II}nd$-Half } & 	$\Lambda CDM$ & 73.06 & 1.05 & 0.32 & 0.02 & ~ & ~ & ~ & ~ & ~ & ~ \\  \cline{3-13}
		& & 	wCDM & 76.24 & 2.11 & 0.37 & 0.04 & -1.54 & 0.3 & ~ & ~ & ~ & ~ \\  \cline{3-13}
		& & 	CPL & 79.03 & 4.29 & 0.36 & 0.06 & -1.27 & 0.46 & 0.78 & 0.86 & ~ & ~ \\  \cline{3-13}
		& & 	JBP & 76.22 & 2.73 & 0.39 & 0.12 & -1.47 & 0.37 & -0.63 & 1.54 & ~ & ~ \\  \cline{3-13}
		& & 	Log & 76.55 & 2.33 & 0.34 & 0.09 & -1.49 & 0.33 & 0.02 & 1.7 & ~ & ~ \\  \cline{3-13}
		& & 	Linear & 79.62 & 5.44 & 0.36 & 0.08 & -1.22 & 0.48 & ~ & ~ & -0.45 & 1.25 \\ \hline
		\end{tabular}
  }
		\caption{Allowed  parameter ranges obtained via MCMC analysis for two equal half bins.}
		\label{MCMC log 2nd half}
\end{table}

\begin{figure}[ht]
\centering
\includegraphics[scale=0.25]{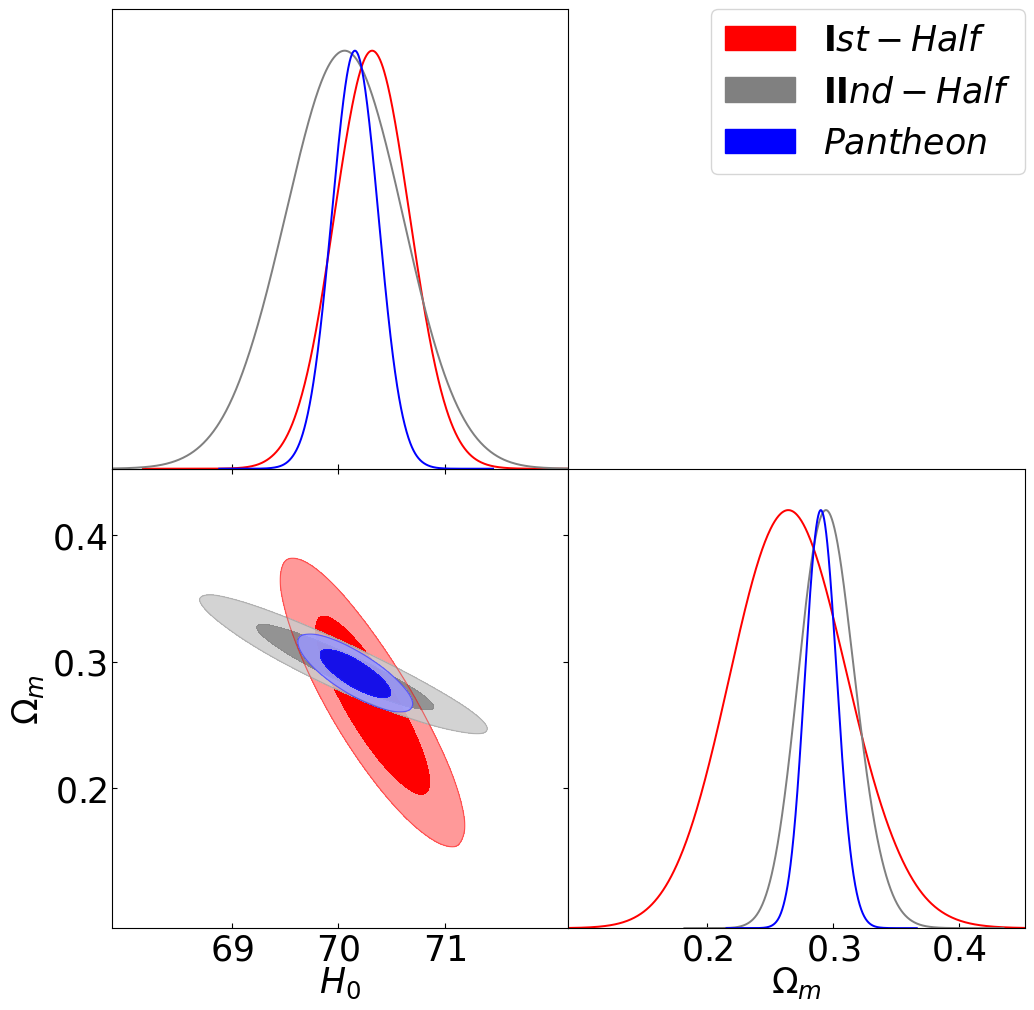} 
\label{Lambda CDM_two bins 0}
\hfill
\includegraphics[scale=0.25]{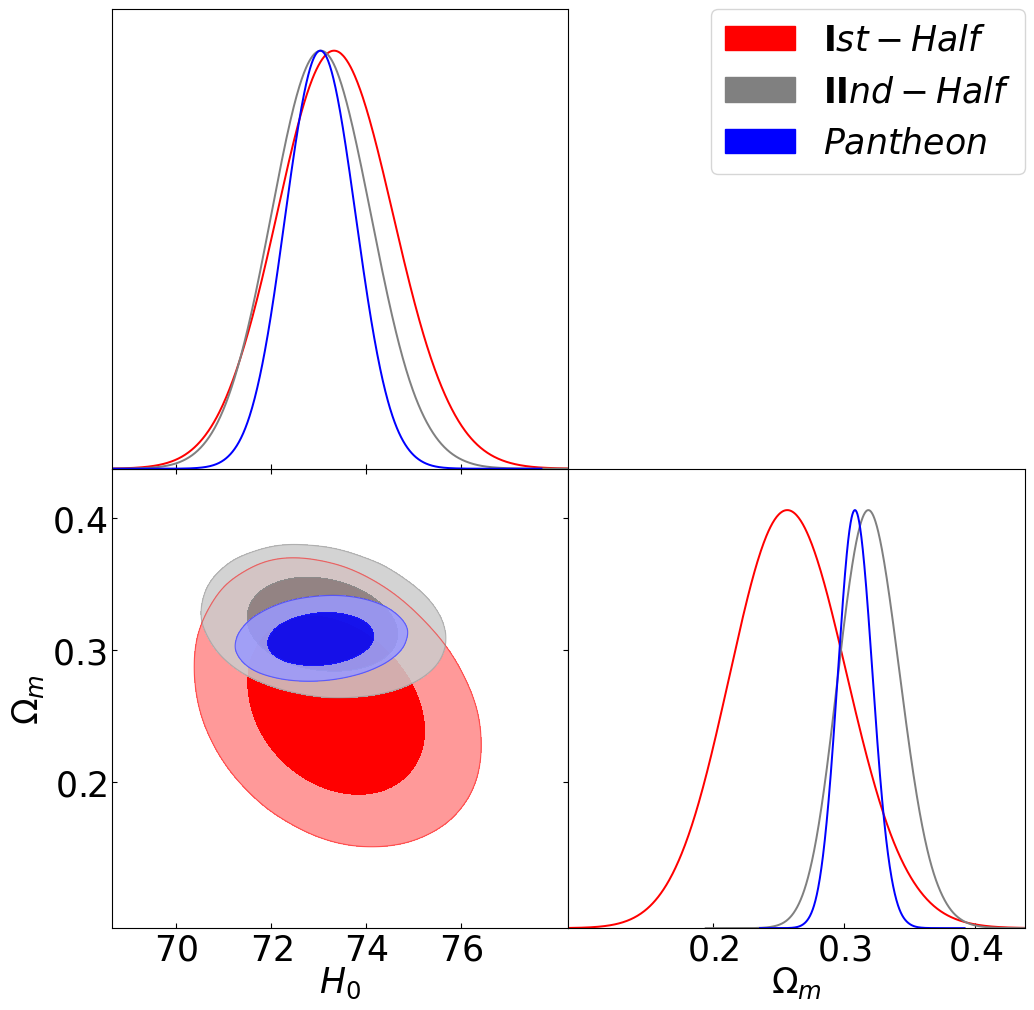} 
\hfill
\includegraphics[scale=0.27]{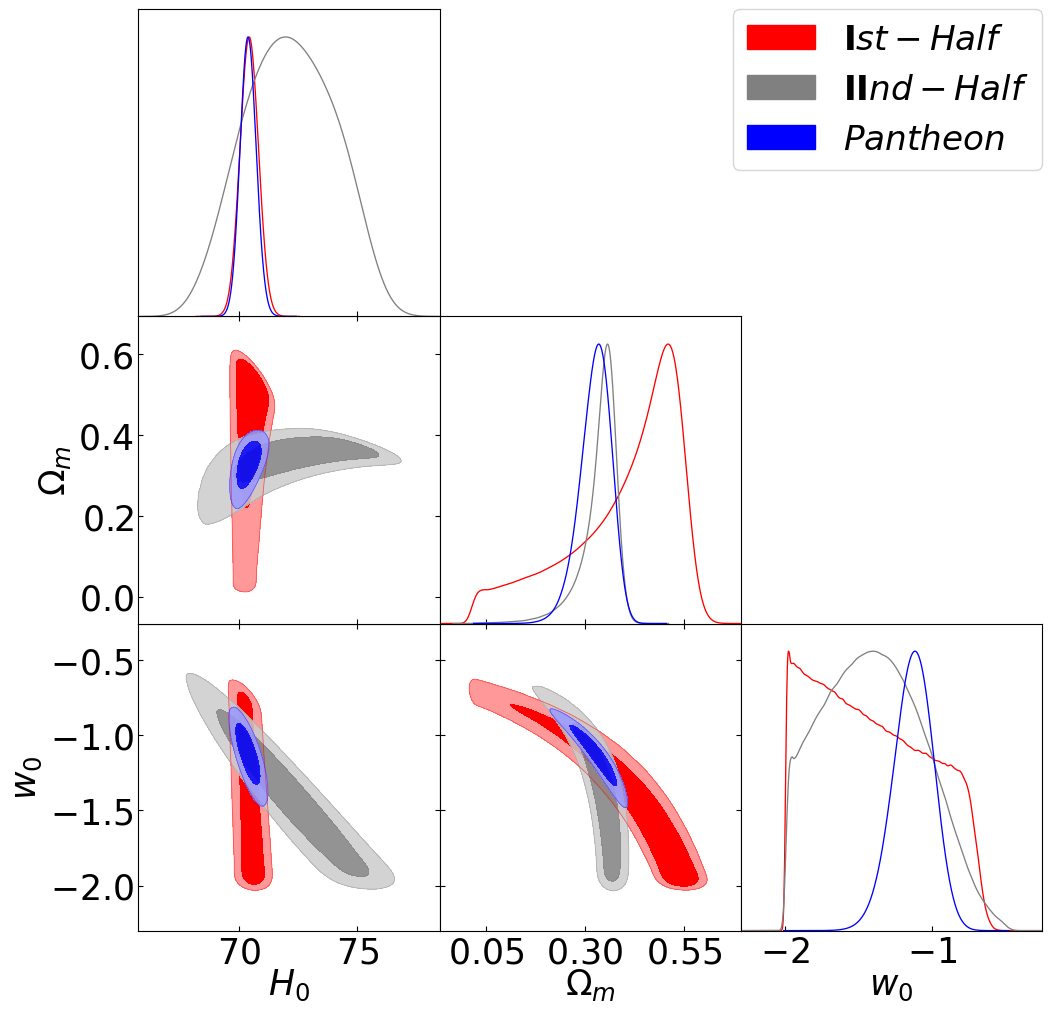} 
\label{wCDM_two bins 0}
\hfill
\includegraphics[scale=0.27]{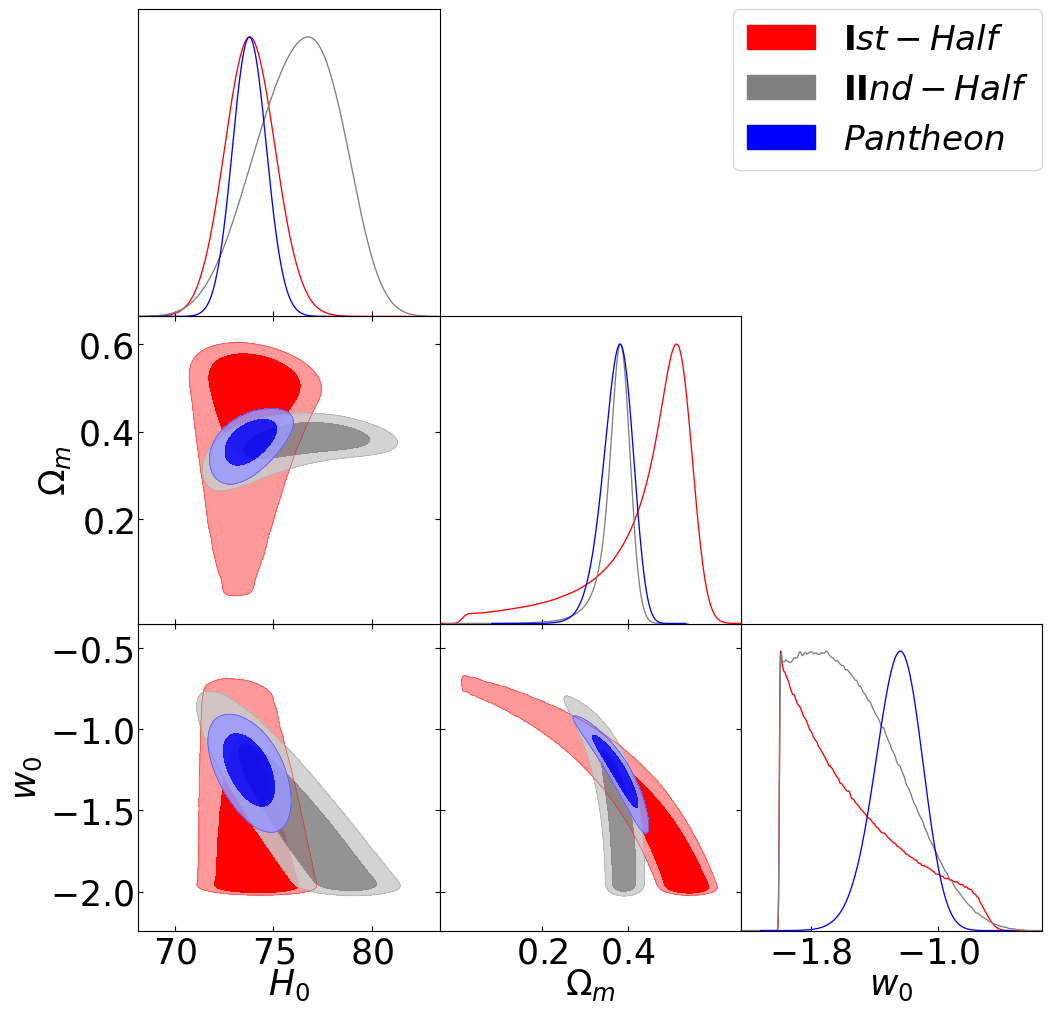} 
\caption{These two equal half bins MCMC $2\sigma$ plots (in the same order as fig \ref{wCDM_0.5-l} ) indicate that both bins contribute equally to the constraints on the full dataset for the Gaussian likelihood, however, it is the second bin which drives the constraints for Logistic likelihood function.
Red contour represents the 1st half bin; grey contour represents the 2nd half bin; blue contour represents the pantheon data. 
}
\label{w CDM half logistic}
\end{figure}

\begin{figure}[ht]
\centering
\includegraphics[scale=0.25]{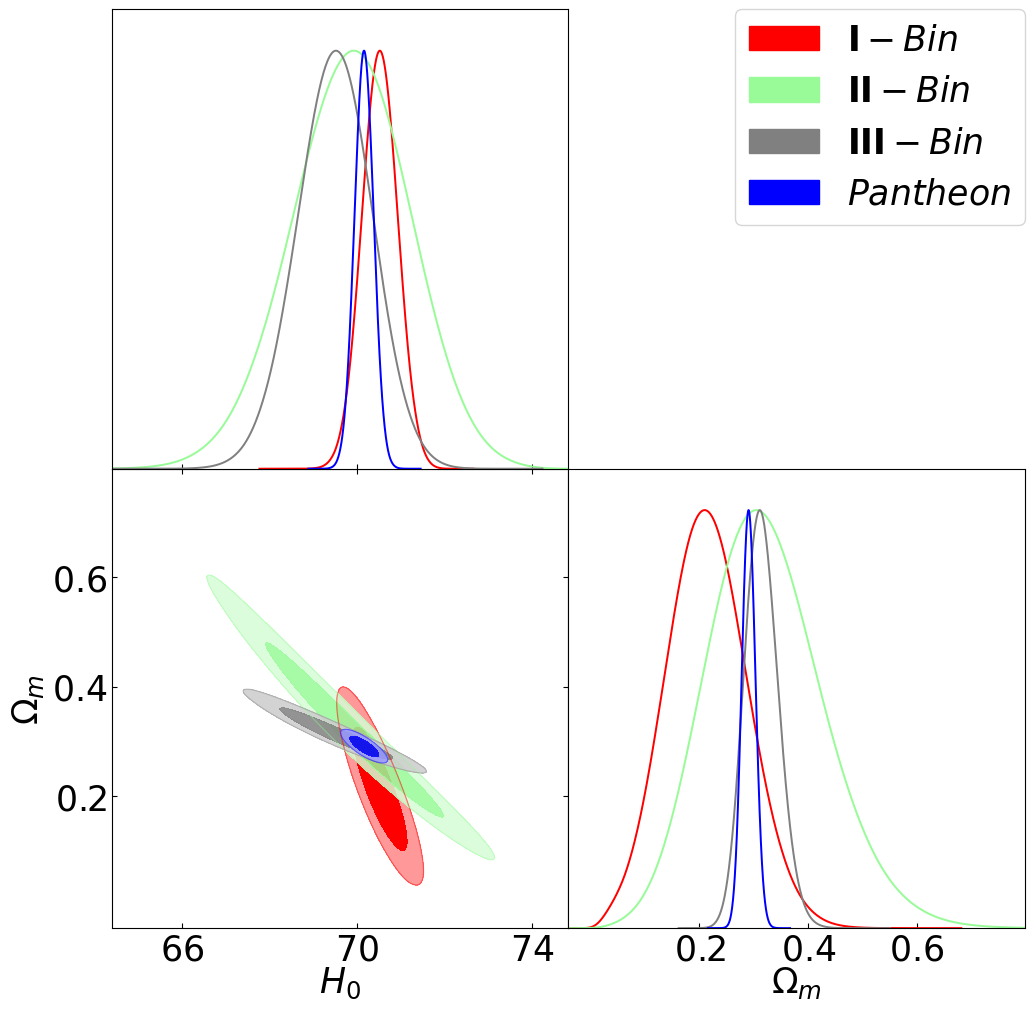} 
\hfill
\includegraphics[scale=0.25]{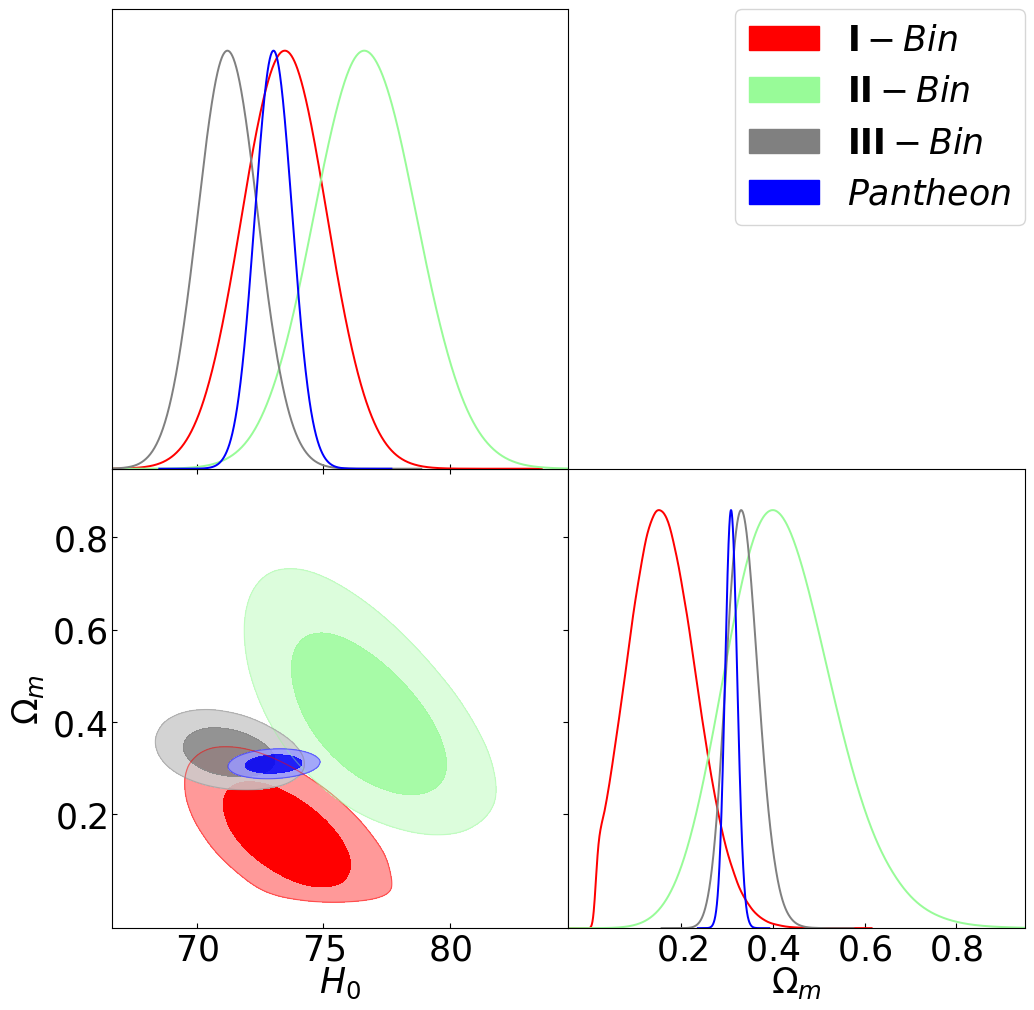} 
\hfill
\includegraphics[scale=0.27]{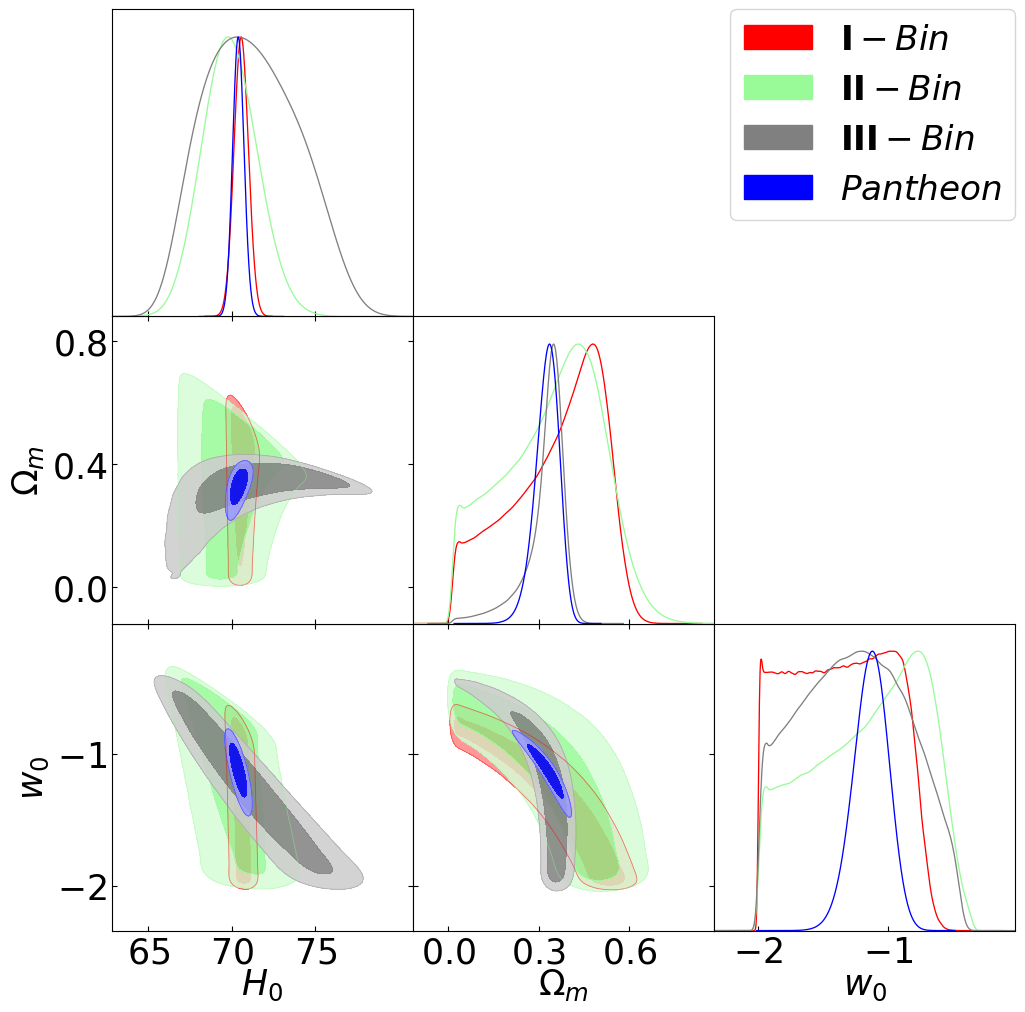} 
\hfill
\includegraphics[scale=0.27]{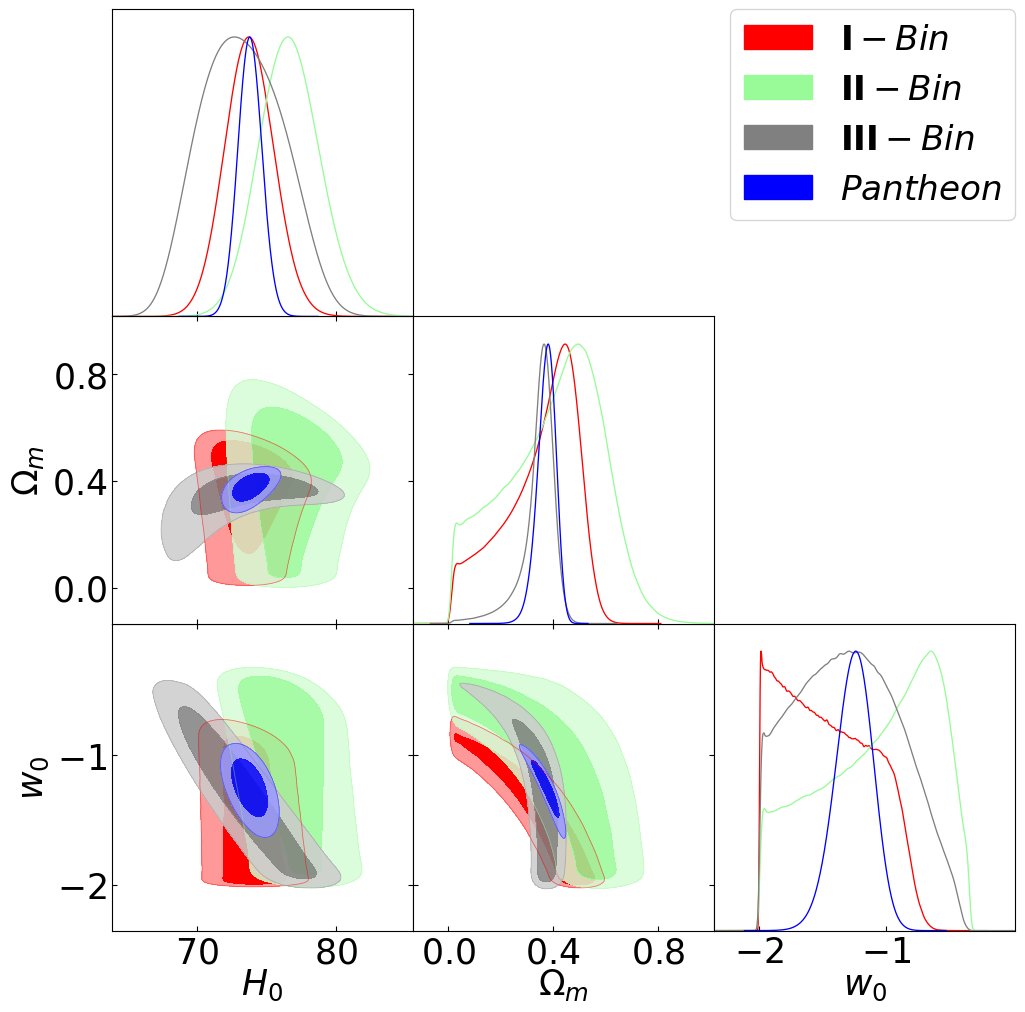} 
\caption{These three equal half bins MCMC $2\sigma$ plots (in the same order as fig \ref{wCDM_0.5-l} ) indicate that all the bins are contributing equally to the constraints on the full dataset for the Gaussian dataset while for Logistic likelihood function, it is the bin 1 and bin 3 which derives the constraints on the full dataset. This indeed is the pattern for other parameterisation too. Red contour represents the 1st bin; pale green contour represents the 2nd bin; grey contour represents the 3rd bin; blue contour represents the pantheon data.}
\label{wCDM_three bins Logistics}
\end{figure}

Employing the third binning scheme, we split the data in three (almost) sections with equal numbers of data points: $349$, $349$, and $350$ with redshift ranges: 0.0101 $<$ z $<$ 0.1769, 0.1771 $<$ z $<$ 0.3374 and 0.3375 $<$ z $<$ 2.26 respectively. The results of MCMC analysis for this are given in table \ref{MCMC log third bin}). It has been suggested that a binned analysis of Pantheon data shows that Hubble constant scales with redshift in the form of power-law as depicted in \cite{montani2024running, Dainotti_2021} Comparing these with the full data set as we have done in the previous two binning schemes, we check  the consistency of the full data set with individual bins (see fig \ref{wCDM_three bins Logistics}) for both the likelihood functions. Here we can see that for logistic likelihood, the 2nd bin (pale green contour) raises the value of the Hubble constant for both the models. The higher redshift data, the 3rd bin (grey contour) is more consistent compared to other two bins for all the parameters. 

\begin{table}[ht]
\centering
\scalebox{0.80}{
    \begin{tabular}{|l|l|l|l|l|l|l|l|l|l|l|l|l|}
    \hline &  &  Model  & $H_0$ & $\sigma_H$  &     $\Omega_{m0}$ &   $\sigma_{\Omega_m} $&     $w_0$ &$\sigma_{w0}$   &  $w^\prime $  & $\sigma_{w^\prime}$    & $w_1$  &$\sigma_{w_1}$  \\
    \cline{3-13}
 \multirow{20}{5em}{ Gaussian}&  \multirow{6}{5em}{ $\textbf{I}st$-Bin} &  $\Lambda CDM $ & 70.514 &  0.406   & 0.214  & 0.074  & &  &   &    
 &       &  \\  \cline{3-13}
 &  &  $w CDM $ &  70.568 & 0.431   &  0.354 &  0.151 & -1.363 & 0.366   &    
 &       & &   \\  
\cline{3-13}
&  &   $CPL$ &  70.580  &  0.437  & 0.351   &  0.153 &  -1.366 & 0.364 & 0.197 &  1.723 &    &  
\\  \cline{3-13}
&  &  $JBP$ &  70.577  &  0.436    & 0.352    &  0.153  &  -1.366 &  0.364  & 0.155    &  1.727  &     &   \\ 
\cline{3-13}
&  &  $log$ &  70.580   &  0.438  &  0.351  &  0.153  &  -1.368  &  0.363  &  0.209   &   1.722   &     &   \\ 
\cline{3-13}
&  &  $linear$ & 70.526 &  0.434 &  0.370   & 0.156  &   -1.354  &    0.374  &    &     &  -0.224   &  1.724 \\  \cline{2-13}

&  \multirow{6}{5em}{ $\textbf{II}nd$-Bin} &  $\Lambda CDM $ & 69.845 &  1.349   & 0.323  & 0.107  & &  &   &    
 &       &  \\ \cline{3-13}
 &  &  $w CDM $ & 69.901 &  1.636   &  0.342 & 0.160  & -1.158 & 0.423 &    
 &       & &   \\  
\cline{3-13}
 &  &  $CPL$ &  69.909  &  1.725 & 0.351  & 0.158   & -1.154  & 0.444     & -0.281  &   1.679 &    &  
\\  \cline{3-13}
&  &  $JBP$ &   69.920   &  1.688  &  0.348  &  0.159  &  -1.158  &   0.439 &  -0.214   &  1.701    &     &   \\ 
\cline{3-13}
&  &  $log$ &  69.906   &  1.739  & 0.354   &  0.158  &  -1.157  & 0.446   &  -0.306   &   1.669   &     &   \\ 
\cline{3-13}
&  &  $linear$ &  69.641   & 1.665  &   0.383    & 0.159  & -1.117  &   0.452 &    &     &  0.221   & 1.676 \\ \cline{2-13}

&   \multirow{6}{5em}{ $\textbf{III}rd$-Bin}  &  $\Lambda CDM $ & 69.483 &  0.856   & 0.315  & 0.031  & &  &   &    
 &       &  \\  \cline{3-13}
&  &   $w CDM $ & 71.218 & 2.745   &  0.319 & 0.069 &  -1.264 & 0.398   &    
 &       & &   \\  
\cline{3-13}
&  &   $CPL$ & 71.591  &  2.987  & 0.317  & 0.089   & -1.264  & 0.410  & -0.169   & 1.657   &    &  
\\ \cline{3-13}
&  &  $JBP$ & 71.343  &  2.815 &  0.320 & 0.077  & -1.259  &  0.412 &   -0.127 &  1.702 &     &   \\ 
\cline{3-13}
&  &  $log$ &  71.566   &  3.183   &  0.326  &  0.091  & -1.258   &  0.421  & -0.369    &   1.544   &     &   \\ 
\cline{3-13}
&  &  $linear$ &  71.066 & 3.175 &  0.357 &  0.082   & -1.192  & 0.420  &    &     &    0.062  & 1.647  \\  
    \cline{1-13}
 \multirow{20}{5em}{ Logistic}&  \multirow{6}{5em}{ $\textbf{I}st$-Bin} &  $\Lambda CDM $ & 73.475 & 1.682 & 0.161 & 0.073  & &  &   &    
 &       &  \\  \cline{3-13}
 &  &  $w CDM $ &  73.754 & 1.766 & 0.347 & 0.137 & -1.456 & 0.344   &    
 &       & &   \\  
\cline{3-13}
&  &   $CPL$ &  74.709  &  1.89  & 0.558   &  0.138 &  -1.219 & 0.485 & 1.709 &  0.936 &    &  
\\  \cline{3-13}
&  &  $JBP$ &  74.08 & 1.82 & 0.304 & 0.147 & -1.505 & 0.309 & 1.005 & 1.571  &     &   \\ 
\cline{3-13}
&  &  $log$ &  73.813 & 1.776 & 0.337 & 0.141 & -1.461 & 0.336 & 0.465 & 1.703   &     &   \\ 
\cline{3-13}
&  &  $linear$ & 75.402 & 2.044 & 0.637 & 0.113 & -1.396 & 0.44 & & & 1.384 & 1.287 \\  \cline{2-13}

&  \multirow{6}{5em}{ $\textbf{II}nd$-Bin} &  $\Lambda CDM $ & 76.674 & 2.025 & 0.421 & 0.118  & &  &   &    
 &       &  \\ \cline{3-13}
 &  &  $w CDM $ & 76.646 & 2.175 & 0.399 & 0.182 & -1.078 & 0.447 &    
 &       & &   \\  
\cline{3-13}
 &  &  $CPL$ &  75.991  &  4.127 & 0.395  & 0.219   & -1.238  & 0.451     & -0.015  &   1.521 &    &  
\\  \cline{3-13}
&  &  $JBP$ &   76.743 & 2.385 & 0.455 & 0.183 & -1.133 & 0.469 & -0.449 & 1.506    &     &   \\ 
\cline{3-13}
&  &  $log$ &  76.685 & 2.25 & 0.418 & 0.175 & -1.1 & 0.457 & -0.252 & 1.658   &     &   \\ 
\cline{3-13}
&  &  $linear$ &  78.787 & 3.68 & 0.513 & 0.173 & -1.199 & 0.475 & & & -0.053 & 1.523 \\ \cline{2-13}

&   \multirow{6}{5em}{ $\textbf{III}rd$-Bin}  &  $\Lambda CDM $ & 71.241 & 1.197 & 0.335 & 0.036  & &  &   &    
 &       &  \\  \cline{3-13}
&  &   $w CDM $ & 73.227 & 2.86 & 0.344 & 0.067 & -1.297 & 0.393   &    
 &       & &   \\  
\cline{3-13}
&  &   $CPL$ & 75.084  &  6.160  & 0.328  & 0.072   & -1.244  & 0.456  & 0.305   & 1.095   &    &  
\\ \cline{3-13}
&  &  $JBP$ & 73.204 & 3.381 & 0.392 & 0.118 & -1.233 & 0.446 & -0.745 & 1.387 &     &   \\ 
\cline{3-13}
&  &  $log$ &  73.55 & 3.272 & 0.341 & 0.094 & -1.269 & 0.415 & -0.323 & 1.582   &     &   \\ 
\cline{3-13}
&  &  $linear$ &  76.921 & 6.651 & 0.329 & 0.09 & -1.193 & 0.475 & & & -0.673 & 1.222  \\  \cline{1-13}
    \end{tabular}
    }
    \caption{Best fit values and the standard deviation in parameters is obtained via mean and standard deviation of the chains obtained from MCMC for Logistic likelihood for three equal bins}
     \label{MCMC log third bin}
\end{table}

While here we have presented the MCMC plots only for the cosmological constant model and constant $w$ model, the overall trend remains consistent across all other  parameterisations. This is true whether we are using the full Pantheon data or binned data in any of the binning scheme mentioned above. For the Logistic likelihood, the contours in parameter space are significantly broader than those with Gaussian likelihood across all parameterisations.

Considering the non-Gaussian behaviour of the residuals across all parameterisations in the Pantheon dataset, we check  which of these parameterisations  is  the most preferred by conducting model selection using Akaike weights, illustrated in Table \ref{Akaike weights}, giving us the relative likelihood of the model being the best fit to the data among a set of competing models.
In our case, we have computed the Akaike weights specifically for models based on the Logistic likelihood function. 
As per the rule of thumb for interpreting AIC test values, the minimum difference between the AIC values of the models should be $2$ \cite{perivolaropoulos2022reanalysis}, so that the model with the lowest AIC value can be considered the best model. 
Hence, a model should at least have a $0.352$ difference in Akaike weight value to be considered better compared to all other models. 
The table \ref{Akaike weights} (with details in the appendix \ref{goodness od fit}) shows us that only some datasets have a single model dominant over all other models. $\Lambda CDM$ and $wCDM$ models show a dominant contribution in most cases. 
All other binned data only have marginally better fitting models, with no significant dominance over others.
Within these, the most significant contribution comes from the models with  non-dynamical dark energy indicating that the higher probability of non-dynamical dark energy model compared to the others in explaining the data which is in agreement with \cite{lovick2023non} in which the analysis is done with Pantheon+ and they showed that a spatially flat $\Lambda CDM$ is preferred by the data over other models.

\begin{table}[H]
    \centering
   \scalebox{0.80}{
    \begin{tabular}{ |c||c||c||c|c||c|c||c|c|c|  }
\hline
 \multirow{2}{4.5em}{Likelihood} & Model & Full data set & \multicolumn{2}{|c||}{$z  \gtrless 0.5$}   & \multicolumn{2}{|c|| }{Two equal bins}   & \multicolumn{3}{|c| }{Three equal bins} \\
 
~ & ~ & ~ & $z<0.5$  & $z>0.5$   & \textbf{I}-Half  & \textbf{II}-Half & \textbf{I} Bin & \textbf{II} Bin & \textbf{III} Bin  \\
\hline

\multirow{6}{4.5em}{Gaussian} & $\Lambda CDM$ & 0.455 & \textbf{0.83} & \textbf{0.39} & \textbf{0.367} & 0.\textbf{578} & \textbf{0.279} & 0.072 & 0.279 \\    
  
~ & $wCDM$ & \textbf{0.472} & 0.054 & 0.125 & 0.191 & 0.243 & 0.223 & 0.008 & 0.077 \\  

~ & $CPL$ & 0.031 & 0.071 & 0.051 & 0.04 & 0.011 & 0.026 & \textbf{0.905} & \textbf{0.531} \\  

~ & $JBP$ & 0.007 & 0.003 & 0.171 & 0.212 & 0.107 & 0.247 & 0.005 & 0.037  \\  

~ & $log$ & 0.008 & 0.031 & 0.187 & 0.179 & 0.06 & 0.225 & 0.01 & 0.076 \\

~ & $linear$ & 0.027 & 0.012 & 0.076 & 0.011 & 0.002 & 0.0 & 0.0 &   0.0 \\  \hline 

\multirow{6}{4.5em}{Logistic} & $\Lambda CDM$ & 0.154 & \textbf{0.301} & 0.214 & 0.02 & 0.153 & 0.132 & \textbf{0.229} & \textbf{0.285} \\    
  
~ & $wCDM$ & \textbf{0.711} & 0.22 & \textbf{0.237} & 0.05 & \textbf{0.549} & 0.179 & 0.206 & 0.274   \\  

~ & $CPL$ & 0.076 & 0.07 & 0.11 & \textbf{0.667} & 0.038 & 0.22 & 0.075 & 0.065 \\  

~ & $JBP$ & 0.023 & 0.015 & 0.23 & 0.217 & 0.154 & \textbf{0.277} & 0.207 & 0.186  \\  

~ & $log$ & 0.015 & 0.097 & 0.209 & 0.047 & 0.105 & 0.191 & 0.21 & 0.189 \\

~ & $linear$ & 0.021 & 0.297 & 0 & 0 & 0 & 0 & 0.073 & 0  \\  \hline 
\end{tabular}
}
    \caption{Akaike weights for Gaussian and Logistic Likelihood. The most dominant contribution is shown in bold face.}
    \label{Akaike weights}
\end{table}

\section{SNLS, SDSS \& PS1 Surveys}
\label{sec 4}
\begin{figure}[ht]
\centering
\includegraphics[scale=0.28]{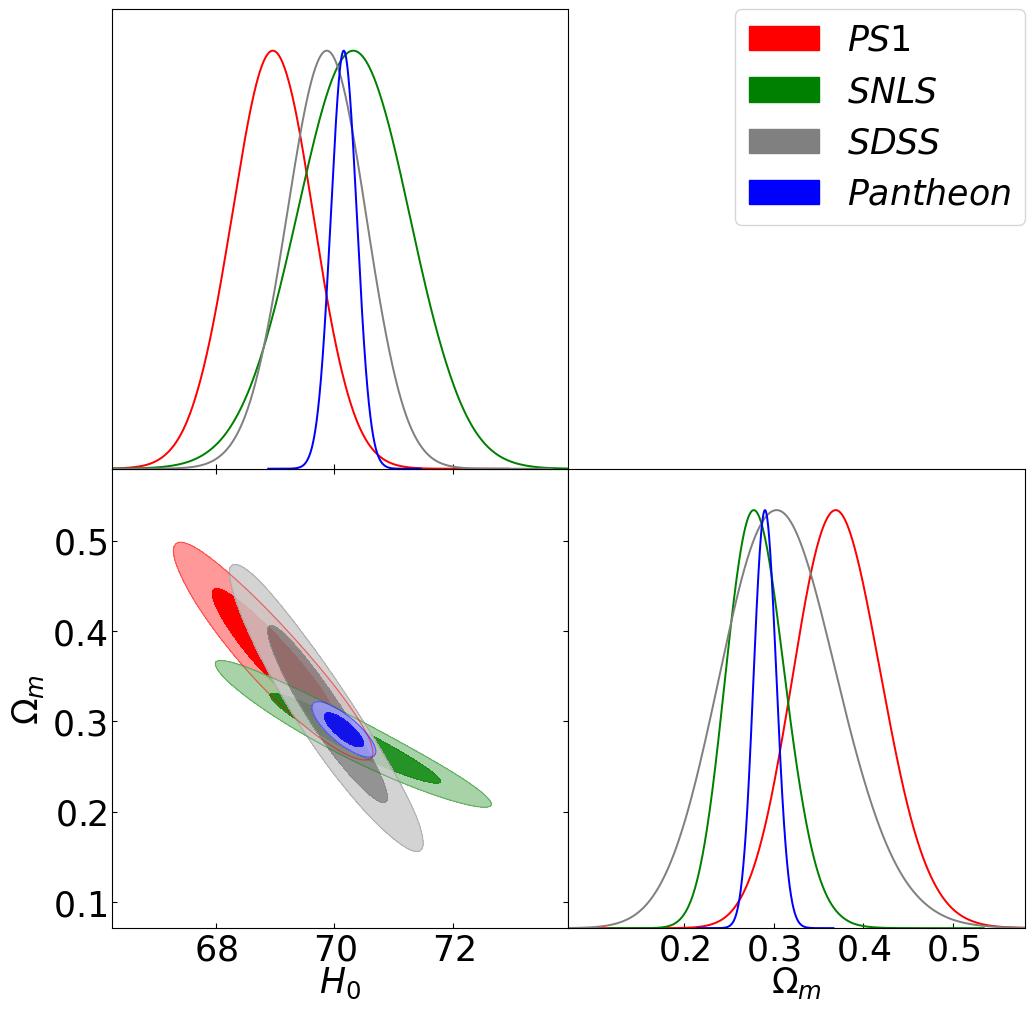} 
\hfill
\includegraphics[scale=0.28]{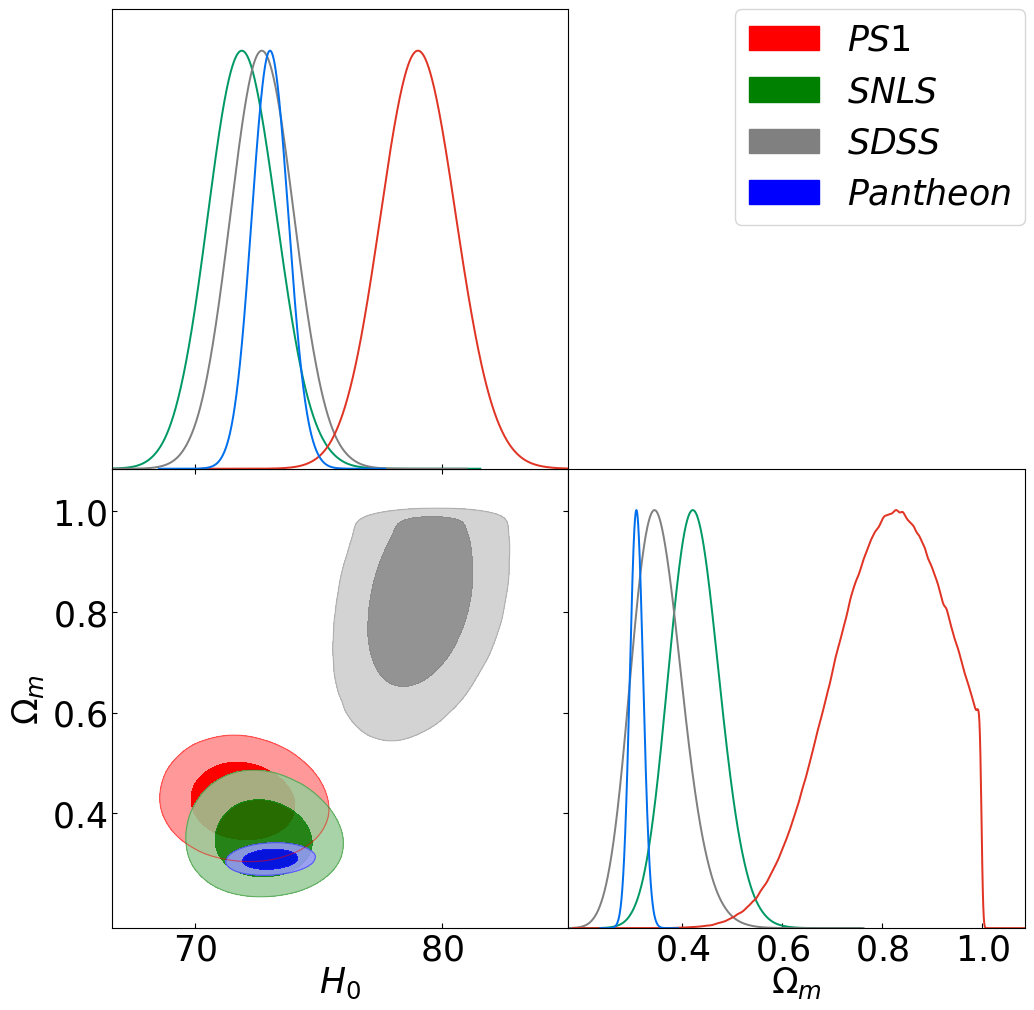} 
\hfill
\includegraphics[scale=0.28]{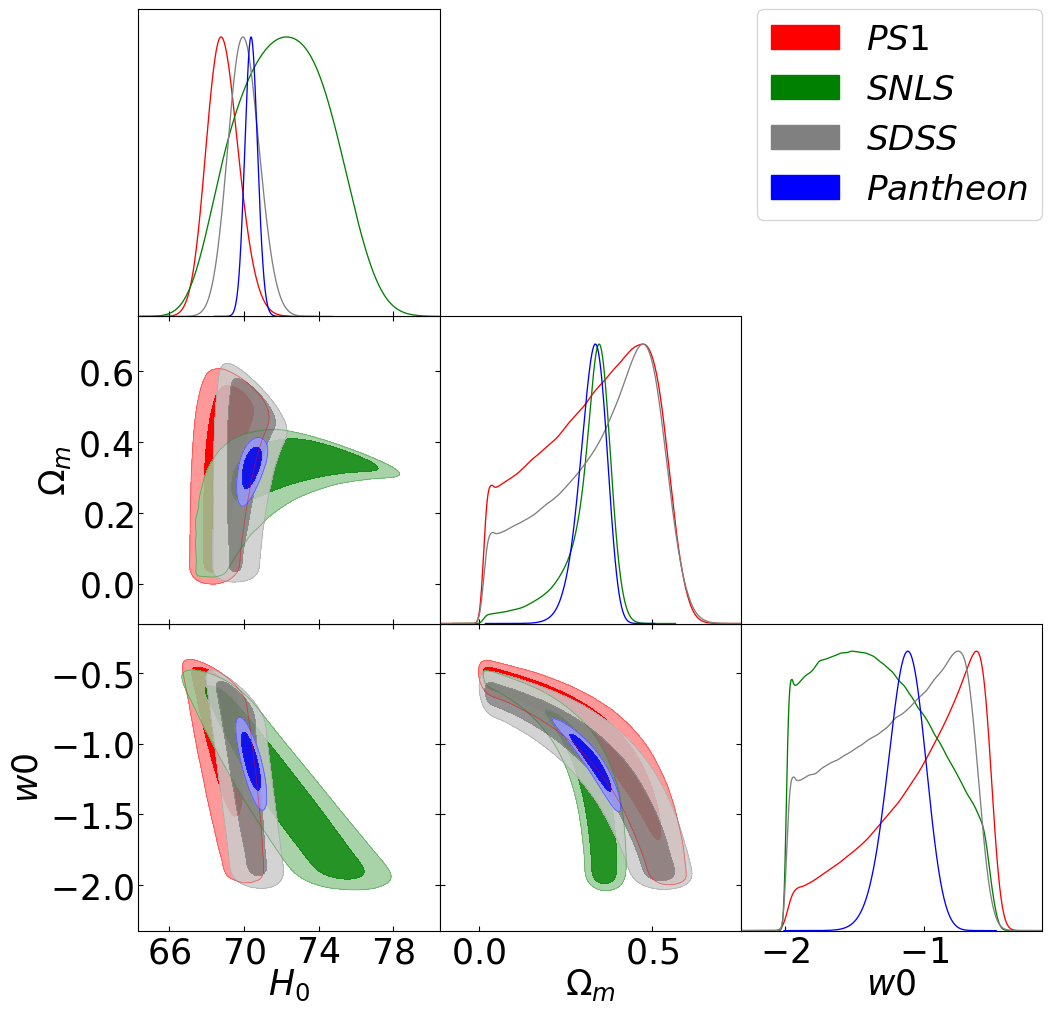} 
\label{wCDM_SS bins gauss}
\hfill
\includegraphics[scale=0.28]{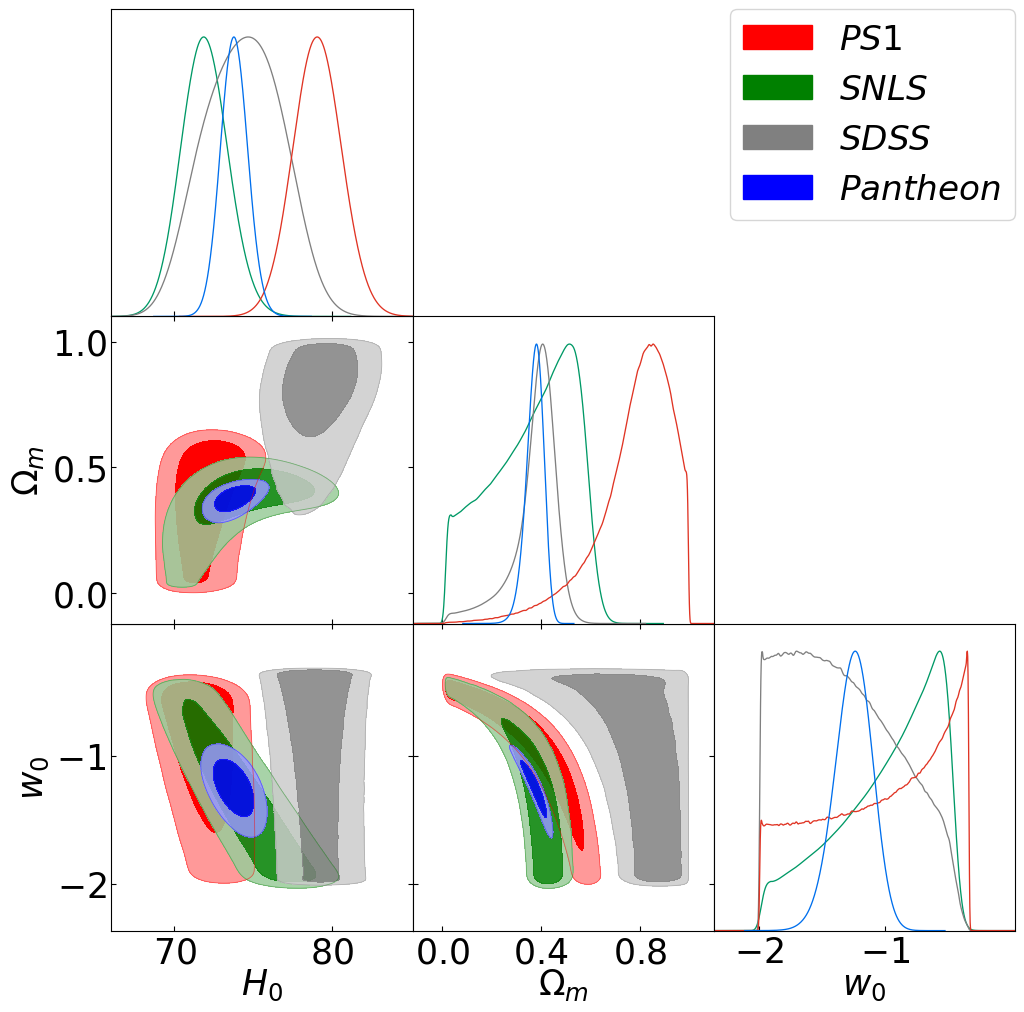} 
\caption{The survey based MCMC $2\sigma$ plots (in the same order as fig \ref{wCDM_0.5-l} ) illustrates the fact that the constraints from the full data are mainly consistent with the SNLS data and choice of the likelihood function can drastically impact the statistics. This also indicates that the different surveys comprising Pantheon data are mutually not consistent. 
The green contour represents SNLS survey; The red contour represents PS1 survey; grey contour represents SDSS survey; blue contour represents the pantheon data. 
Note that in case of Gaussian likelihood, SNLS part of the Pantheon dataset give rise to the larger error bars on $H_0$ compared to SDSS and PS1 irrespective of the parameterisations. 
A similar claim can be made for Logistic likelihood except in the case of $\Lambda CDM$. }
\label{wCDM_SS bins log}
\end{figure}

The Pantheon dataset compiles data from surveys including SDSS, SNLS, and PS1 \cite{2018ApJ...859..101S}. SNe from different surveys of Pantheon are cross-calibrated \cite{scolnic2015supercal}.
The $236$ SNLS data points, $335$ SDSS data points and $279$ PS1 data points in the Pantheon dataset are examined here.
The Supernova Legacy Survey (SNLS) \cite{conley2010supernova,Sullivan_2011} measures the luminosity distances of Type Ia supernovae, targeting those with redshifts between 0.2 and 0.9.
The  SDSS (\cite{frieman2007sloan}, \cite{kessler2009first}) sample targets a redshift range of $0.05$ to $0.4$.
Whereas the Pan-Starrs \cite{chambers2016pan} survey in Pantheon spans the redshift range $0.026<z<0.631$. 
With these SNLS, SDSS and PS1 subsets of Pantheon, we perform the MCMC analysis whose results
 are as given in table \ref{mCmC logistic for SDSS}. From the MCMC plots \ref{wCDM_SS bins log} for cosmological constant model and constant $w_0$ parameterisation, we can see that, in case of logistic likelihood, the SDSS data is inconsistent with the pantheon data and shows around 4$\sigma$ deviation. The higher redshift data, that is the SNLS survey data is consistent with the full dataset. 
 Similar results are discussed in \cite{ghosh2024consistencytestssdssdesi}, where Baryon Acoustic Oscillations (BAO) data from SDSS and DESI were used to reconstruct Hubble parameter $H(z)$ and deceleration parameter $q(z)$, from both datasets using a non-parametric reconstruction method. 
 They have found that the reconstructed parameters from SDSS and DESI are significantly inconsistent and are only marginally consistent with the cosmological constant model within 3$\sigma$ confidence level. 
The method used for cross-calibrating different surveys in the Pantheon dataset appears to implicitly assume a Gaussian likelihood function. However, this assumption is not appropriate as the residuals do not follow Gaussian behaviour, raising concerns about the possibility of underestimation of errors from the Gaussian likelihood.

\begin{table}[t]
\centering
\scalebox{0.8}{
    \begin{tabular}{|l|l|l|l|l|l|l|l|l|l|l|l|l|}
    \hline
      &  &   Models & $H_0$ & $\sigma_H$ & $\Omega_m$ & $\sigma_{\Omega_m}$ & $w_0$ & $\sigma_{w_0}$ & $w'$ & $\sigma_{w'}$ & $w_1$ & $\sigma_{w_1}$ \\
       \cline{3-13} 
\multirow{12}{5em}{Gaussian}  &  \multirow{5}{5em}{SNLS}   &    $\Lambda CDM$ & 70.300 & 0.951  &0.282  & 0.033  & ~ & ~ & ~ & ~ & ~ & ~ \\ \cline{3-13} 
  &      &   wCDM &72.159   & 2.380  & 0.310  & 0.077  & -1.344  & 0.393  & ~ & ~ & ~ & ~ \\ \cline{3-13}
  &     &    CPL & 74.457  &  4.476 & 0.319  & 0.092  &  -1.211  & 0.466  &  0.568  &  0.924  & ~ & ~ \\\cline{3-13} 
  &     &    JBP & 72.431 & 2.832  &  0.343  &     0.133  &  -1.349 &  0.411 &  -0.527 & 1.500  & ~ & ~ \\\cline{3-13} 
 &       &   Log & 72.763 &  2.563 & 0.309 &    0.093  & -1.385  & 0.376  &  -0.139 & 1.619  & ~ & ~ \\ \cline{3-13} 
 &       &   Linear & 74.548  &  4.405  &  0.325  &  0.089 &  -1.189  & 0.478  & ~ & ~ &  -0.658  & 1.147   \\ 

    \cline{2-13}
    
   &   \multirow{5}{5em}{SDSS}   &   $\Lambda CDM$ & 69.857 & 0.668 & 0.308  & 0.065  & ~ & ~ & ~ & ~ & ~ & ~ \\ \cline{3-13} 
   &      &  wCDM & 70.012 & 0.827  & 0.346  & 0.152   & -1.211  & 0.403  & ~ & ~ & ~ & ~ \\\cline{3-13} 
   &      &  CPL  & 70.604  & 1.440  & 0.441  &  0.169  &   -1.184  &  0.470  &  0.687   & 1.095  & ~ & ~ \\ \cline{3-13} 
  &       &  JBP & 70.043  & 1.115  & 0.382   & 0.161   & -1.244  &  0.436  & -0.349  & 1.536  & ~ & ~ \\ \cline{3-13} 
  &       &  Log &  70.025 & 0.936   & 0.354   & 0.151  & -1.222  & 0.419  &  -0.141 & 1.679   & ~ & ~ \\ \cline{3-13}
   &     &   Linear & 70.953  & 1.479   &  0.468  &  0.167 & -1.183  &  0.477 & ~ & ~ &  -0.220   & 1.409  \\  
   
      \cline{2-13}
    
   &   \multirow{5}{5em}{PS1}   &   $\Lambda CDM$ & 68.952  & 0.688 & 0.373   &  0.049  & ~ & ~ & ~ & ~ & ~ & ~ \\ \cline{3-13} 
   &      &  wCDM &  68.902 &  0.866  & 0.323  &  0.154  & -1.005   & 0.383   & ~ & ~ & ~ & ~ \\\cline{3-13} 
   &      &  CPL  & 68.821   &  1.015  &  0.356  &  0.145  & -1.004   &  0.407    &  -0.575   &  1.536  & ~ & ~ \\ \cline{3-13} 
  &       &  JBP &  68.852 & 0.977   &  0.343  &  0.147  & -0.997   &   0.403  &  -0.448 &  1.614  & ~ & ~ \\ \cline{3-13} 
  &       &  Log &  68.797  &   1.033  & 0.364  &  0.143  &  -1.007  & 0.408   & -0.649  &  1.487   & ~ & ~ \\ \cline{3-13}
   &     &   Linear &  68.660  & 0.999    & 0.398    &  0.142  &  -0.997  & 0.399   & ~ & ~ &   0.406  &  1.567 \\

   \hline
   
	\multirow{12}{5em}{Logistic} &   \multirow{5}{5em}{SNLS} & 		$\Lambda CDM$ & 72.74 & 1.30 & 0.35 & 0.05 & ~ & ~ & ~ & ~ & ~ & ~ \\\cline{3-13} \cline{3-13} 
	& & 		wCDM & 74.38 & 2.41 & 0.37 & 0.09 & -1.33 & 0.42 & ~ & ~ & ~ & ~ \\ \cline{3-13} \cline{3-13} 
	& & 		CPL & 77.52 & 4.49 & 0.39 & 0.08 & -1.22 & 0.47 & 0.78 & 0.96 & ~ & ~ \\\cline{3-13} \cline{3-13} 
	& & 		JBP & 75.04 & 2.83 & 0.38 & 0.15 & -1.38 & 0.4 & -0.34 & 1.55 & ~ & ~ \\ \cline{3-13} \cline{3-13} 
	& & 		Log & 75.10 & 2.50 & 0.37 & 0.11 & -1.39 & 0.39 & -0.02 & 1.64 & ~ & ~ \\ \cline{3-13} \cline{3-13} 
	& & 		Linear & 78.92 & 5.85 & 0.40 & 0.09 & -1.2 & 0.47 & ~ & ~ & -0.26 & 1.33 \\ \cline{3-13} \cline{2-13}

	&  \multirow{5}{5em}{SDSS} & 		$\Lambda CDM$ & 79.06 & 1.53 & 0.81 & 0.11 & ~ & ~ & ~ & ~ & ~ & ~ \\\cline{3-13} 
	&  & 		wCDM & 79.09 & 1.54 & 0.78 & 0.16 & -1.04 & 0.49 & ~ & ~ & ~ & ~ \\\cline{3-13} 
	& & 		CPL & 79.38 & 1.58 & 0.79 & 0.17 & -1.17 & 0.49 & 0.46 & 1.41 & ~ & ~ \\ \cline{3-13} 
	& & 		JBP & 79.76 & 1.74 & 0.68 & 0.25 & -1.05 & 0.46 & 0.9 & 1.59 & ~ & ~ \\\cline{3-13} 
	& & 		Log & 79.25 & 1.56 & 0.74 & 0.21 & -1.02 & 0.49 & 0.48 & 1.72 & ~ & ~ \\ \cline{3-13} 
	& & 		Linear & 79.75 & 1.65 & 0.80 & 0.19 & -1.23 & 0.47 & ~ & ~ & 0.11 & 1.73 \\ 
 
     \cline{2-13}
    
   &   \multirow{5}{5em}{PS1}   &   $\Lambda CDM$ & 71.93  & 1.4  & 0.43   & 0.05   & ~ & ~ & ~ & ~ & ~ & ~ \\ \cline{3-13} 
   &      &  wCDM   & 71.93 & 1.48 & 0.36 & 0.16 & -0.99 & 0.4 & ~ & ~ & ~ & ~ \\ \cline{3-13} 
   &      &  CPL    & 71.92 & 1.57 & 0.40 & 0.15 & -1.02 & 0.42 & -0.43 & 1.56 & ~ & ~ \\ \cline{3-13} 
   &      &  JBP    & 71.93 & 1.55 & 0.39 & 0.16 & -1.01 & 0.42 & -0.3 & 1.63 & ~ & ~ \\ \cline{3-13} 
   &      &  Log    & 71.89 & 1.59 & 0.41 & 0.15 & -1.03 & 0.42 & -0.5 & 1.52 & ~ & ~ \\ \cline{3-13}
   &      &  Linear & 71.75 & 1.56 & 0.44 & 0.15 & -1.02 & 0.41 & ~ & ~ & 0.27 & 1.59 \\

 \hline
		\end{tabular}
  }
		\caption{Best fit values and the standard deviation in parameters is obtained via mean and standard deviation of the chains obtained from MCMC for SDSS, SNLS and PS1}
		\label{mCmC logistic for SDSS}
\end{table}

\begin{figure}[ht]
\centering
\includegraphics[scale=0.55]{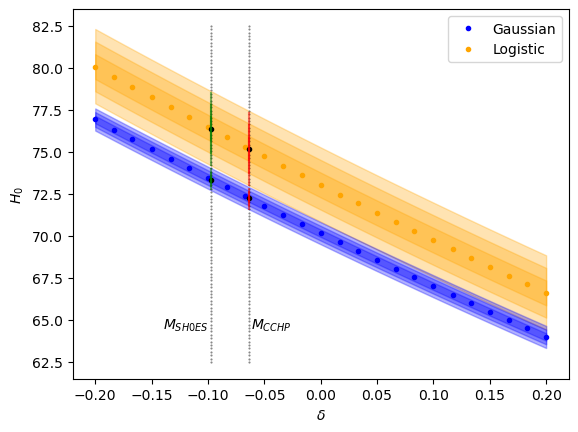} 
\caption{Deviation in the value of observed Distance Modulus $\delta$ from the Pantheon dataset value as a function of the  corresponding estimated value of Hubble constant $H_0$. The blue dots represents value estimated using Gaussian Likelihood; The orange dots represents value estimated using Logistic Likelihood; The shaded bands corresponds to 1$\sigma$, 2$\sigma$, and 3$\sigma$ confidence ranges; The black dots denote the estimated value, while the red and green error bars represent the error ranges of $M_{CCHP}$ and $M_{SH0ES}$ respectively.}
\label{MUvsH}
\end{figure}

We further investigate the sensitivity of the Hubble constant to small variations in the value of distance modulus. Using $\Lambda CDM$ model, we  estimate the value of $H_0$ using Likelihood maximisation for Gaussian and Logistic likelihoods. 
As shown in fig \ref{MUvsH}, even minor deviations in the value of $\mu$ results in significant shifts in the estimated value of $H_0$. This suggests that even a small change in the value of distance modulus can have a significant effect on the Hubble tension indicating the need to calibrate the data more precisely.

\FloatBarrier
\section{Summary \& Conclusion}
\label{sec 5}
In this work, we presented a detailed analysis of Pantheon compilation of supernova type Ia dataset,  considering both the full dataset, binned data  on the basis of redshift and the constituent surveys. 
We explored the deviation from Gaussian distribution of residuals in the distance modulus.
Different parameterisations for equation of state parameter of dark energy were considered  and  various statistical goodness of fit tests verified that Logistic likelihood function fits better to the Pantheon data compared to the Gaussian likelihood.
This is true for the full dataset and lower redshift bins whereas for the higher redshift bins, the comparison is inconclusive. 
The Markov Chain Monte Carlo method for parameter estimation using the Logistic likelihood for the full dataset suggests a value of the Hubble constant closer to that obtained from late-time measurements with larger error bars.
Therefore choice of statistics can provide a way in reducing Hubble tension, in this case at the cost of increased uncertainties. 
This holds true irrespective of the dark energy parametrisations chosen for the analysis. 
In the data subsets analysis, it has been noted that higher redshift bins yield lower values of the Hubble constant when compared to lower redshift bins. This result is consistent with the results from \cite{colgain2024putting, o2022revealing, krishnan2020there}.
This observation is consistent across the cosmological constant model.
Our results for the cosmological constant are consistent with the ones obtained in \cite{DAINOTTI202430} and we have extended the analysis to dynamical dark energy models and different subsets of the data. 

Additionally, under the Gaussain/Logistic likelihood scenario, with the few exception of certain bins, we found a preference for constant $w$ dark energy models over the others.
Binned datasets with non-overlapping redshift ranges are found to be consistent with the full dataset, yielding parameter estimates that agree within the $2\sigma$ range. In contrast, for binned data with overlapping redshift ranges, where different surveys are analysed, the consistency is influenced by the choice of the likelihood function. 
Even though Logistic likelihood fits the data better compared to Gaussian likelihood, we have observed inconsistencies between surveys in Logistic likelihood, whereas Gaussian likelihood yields consistent results. This indicates that the Pantheon dataset implicitly assumes Gaussian distribution in the residuals during the compilation of surveys, which may lead to underestimation of parameter ranges allowed. This naturally raises the question of how Gaussian behaviour of residuals can be restored and, once achieved, how it impacts parameter estimation, given its importance in the context of the Central Limit Theorem.

The work also shows that even a marginal shift in the absolute magnitude can change the value of the Hubble constant significantly, indicating that there may be unresolved systematics in the data. 
Moreover, the results indicate that a better statistical modelling merits further study.

\acknowledgments
Ramanpreet Singh thanks the fellowship support from   University Grant Commission, India via fellowship  NTA ref. No. 211610173227. 
.We acknowledge IISER Mohali for providing computing resources of High Performance Computing Facility at IISER Mohali.
%We would like to thank the support provided by the High-Performance Computing facility at IISER Mohali, which was crucial for majority of the computations performed in this work. 
The authors would also like to thank J. S. Bagla and Anshul Srivastava for useful discussions.  HKJ thanks
NCRA-TIFR for hospitality, as this manuscript was completed during a sabbatical from IISER Mohali.
This research has made use of NASA’s Astrophysics Data System

\appendix
\label{appendix}
\section{Goodness of fit}
\label{goodness od fit}
We have provided the values for the Akaike Information Criterion (AIC), Bayesian Information Criterion (BIC), and Kolmogorov-Smirnov (KS) test for each of the parameterisations applied to both the full and binned datasets. These metrics are used to evaluate and compare the models performance, with AIC and BIC serving as criteria for model selection based on goodness-of-fit and complexity, and the KS test evaluates the distributional differences between the observed and expected data.

\begin{table}[H]
\centering
\scalebox{0.7}{
\begin{tabular}{|l|l|l|l|l|l|}
\hline
    \multicolumn{6}{|c|}{Pantheon Data} \\ \hline 
Estimated using & Model & Null hypothesis & AIC & BIC & KS-test \\ \hline
\multirow{10}{*}{Gaussian Likelihood} 
& \multirow{2}{*}{LCDM}   & Gaussian  & 2976.06 & 2985.97 & 0.25 \\ \cline{3-6}
&        & Logistic  & 2964.21 & 2974.12 & 0.99 \\ \cline{2-6}
& \multirow{2}{*}{wCDM}   & Gaussian  & 2975.99 & 2985.90 & 0.17 \\ \cline{3-6}
&        & Logistic  & 2964.21 & 2974.12 & 0.95 \\ \cline{2-6}
& \multirow{2}{*}{CPL}  & Gaussian  & 2981.46 & 2991.37 & 0.16 \\ \cline{3-6}
&        & Logistic  & 2969.59 & 2979.50 & 0.98 \\ \cline{2-6}
& \multirow{2}{*}{JBP}   & Gaussian  & 2984.38 & 2994.29 & 0.19 \\ \cline{3-6}
&        & Logistic  & 2972.77 & 2982.68 & 0.97 \\ \cline{2-6}
& \multirow{2}{*}{Log}   & Gaussian  & 2984.06 & 2993.96 & 0.17 \\ \cline{3-6}
&        & Logistic  & 2972.32 & 2982.23 & 0.98 \\ \cline{2-6}
& \multirow{2}{*}{Linear} & Gaussian  & 2981.71 & 2991.62 & 0.16 \\ \cline{3-6}
&        & Logistic  & 2969.91 & 2979.81 & 0.97 \\ \hline

\multirow{10}{*}{Logistic Likelihood}
& \multirow{2}{*}{LCDM}   & Gaussian  & 2962.91 & 2972.82 & 0.16 \\ \cline{3-6}
&        & Logistic  & 2947.96 & 2957.87 & 0.98 \\ \cline{2-6}
& \multirow{2}{*}{wCDM}   & Gaussian  & 2959.94 & 2969.85 & 0.10 \\ \cline{3-6}
&        & Logistic  & 2944.90 & 2954.81 & 0.98 \\ \cline{2-6}
& \multirow{2}{*}{CPL}  & Gaussian  & 2964.43 & 2974.34 & 0.14 \\ \cline{3-6}
&        & Logistic  & 2949.39 & 2959.29 & 0.81 \\ \cline{2-6}
& \multirow{2}{*}{JBP}   & Gaussian  & 2966.53 & 2976.44 & 0.12 \\ \cline{3-6}
&        & Logistic  & 2951.77 & 2961.68 & 0.66 \\ \cline{2-6}
& \multirow{2}{*}{Log}   & Gaussian  & 2967.38 & 2977.29 & 0.14 \\ \cline{3-6}
&        & Logistic  & 2952.61 & 2962.52 & 0.71 \\ \cline{2-6}
& \multirow{2}{*}{Linear} & Gaussian  & 2966.76 & 2976.67 & 0.13 \\ \cline{3-6}
&        & Logistic  & 2951.95 & 2961.86 & 0.74 \\ \hline
\end{tabular}}
\caption{Goodness of fit test values for the Pantheon data}
\label{goftfd}
\end{table}

\begin{table}[H]
\centering
\scalebox{0.8}{
\begin{tabular}{|l|l|l|l|l|l|l|l|l|}
\hline
    \multicolumn{9}{|c|}{Redshift Bins} \\ \hline
  \multicolumn{3}{|c|}{}  & \multicolumn{3}{|c|}{$z < 0.5$} & \multicolumn{3}{|c|}{$z > 0.5$}\\ \hline 
Estimated using & Model & Null hypothesis & AIC & BIC & KS-test & AIC & BIC & KS-test \\ \hline
\multirow{10}{*}{Gaussian Likelihood} 
& \multirow{2}{*}{$\Lambda$CDM}   & Gaussian  & 2398.77 & 2408.22 & 0.20 & 576.11 & 582.86 & 0.86 \\ \cline{3-9}
&        & Logistic  & 2388.49 & 2397.94 & 0.98 & 576.68 & 583.43 & 0.93 \\ \cline{2-9}
& \multirow{2}{*}{wCDM}   & Gaussian  & 2404.23 & 2413.68 & 0.23 & 578.39 & 585.14 & 0.68 \\ \cline{3-9}
&        & Logistic  & 2394.36 & 2403.81 & 1.00 & 578.65 & 585.40 & 0.90 \\ \cline{2-9}
& \multirow{2}{*}{CPL}  & Gaussian  & 2403.70 & 2413.15 & 0.22 & 580.19 & 586.94 & 0.47 \\ \cline{3-9}
&        & Logistic  & 2393.62 & 2403.07 & 0.99 & 579.94 & 586.69 & 0.67 \\ \cline{2-9}
& \multirow{2}{*}{JBP}   & Gaussian  & 2410.06 & 2419.51 & 0.37 & 577.76 & 584.51 & 0.63 \\ \cline{3-9}
&        & Logistic  & 2400.67 & 2410.11 & 1.00 & 577.71 & 584.46 & 0.84 \\ \cline{2-9}
& \multirow{2}{*}{Log}   & Gaussian  & 2405.37 & 2414.82 & 0.28 & 577.58 & 584.33 & 0.78 \\ \cline{3-9}
&        & Logistic  & 2395.52 & 2404.96 & 1.00 & 577.53 & 584.28 & 0.88 \\ \cline{2-9}
& \multirow{2}{*}{Linear} & Gaussian  & 2407.24 & 2416.69 & 0.27 & 579.38 & 586.13 & 0.47 \\ \cline{3-9}
&        & Logistic  & 2397.59 & 2407.03 & 0.99 & 579.30 & 586.05 & 0.70 \\ \hline

\multirow{10}{*}{Logistic Likelihood}
& \multirow{2}{*}{$\Lambda$CDM}   & Gaussian  & 2386.64 & 2396.09 & 0.21 & 573.33 & 580.08 & 0.79 \\ \cline{3-9}
&        & Logistic  & 2373.92 & 2383.37 & 0.96 & 573.44 & 580.19 & 0.99 \\ \cline{2-9}
& \multirow{2}{*}{wCDM}   & Gaussian  & 2387.57 & 2397.02 & 0.21 & 573.45 & 580.20 & 0.89 \\ \cline{3-9}
&        & Logistic  & 2374.55 & 2384.00 & 0.83 & 573.24 & 579.99 & 0.94 \\ \cline{2-9}
& \multirow{2}{*}{CPL}  & Gaussian  & 2389.91 & 2399.36 & 0.27 & 575.55 & 582.30 & 0.80 \\ \cline{3-9}
&        & Logistic  & 2376.85 & 2386.30 & 0.91 & 574.77 & 581.52 & 1.00 \\ \cline{2-9}
& \multirow{2}{*}{JBP}   & Gaussian  & 2392.36 & 2401.81 & 0.23 & 573.86 & 580.61 & 0.86 \\ \cline{3-9}
&        & Logistic  & 2379.87 & 2389.31 & 0.86 & 573.30 & 580.05 & 0.99 \\ \cline{2-9}
& \multirow{2}{*}{Log}   & Gaussian  & 2389.09 & 2398.54 & 0.21 & 574.07 & 580.82 & 0.86 \\ \cline{3-9}
&        & Logistic  & 2376.18 & 2385.63 & 0.82 & 573.49 & 580.24 & 0.99 \\ \cline{2-9}
& \multirow{2}{*}{Linear} & Gaussian  & 2392.29 & 2401.74 & 0.21 & 574.10 & 580.85 & 0.92 \\ \cline{3-9}
&        & Logistic  & 2379.75 & 2389.20 & 0.83 & 573.58 & 580.33 & 0.96 \\ \hline
\end{tabular}}
\caption{Goodness of fit test values for $z \gtrless 0.5$ Bins}
\label{goft5b}
\end{table}

\begin{table}[H]
\centering
\scalebox{0.8}{
\begin{tabular}{|l|l|l|l|l|l|l|l|l|}
\hline
    \multicolumn{9}{|c|}{Two Equal Bins} \\ \hline
  \multicolumn{3}{|c|}{}  & \multicolumn{3}{|c|}{First Bin} & \multicolumn{3}{|c|}{Second Bin}\\ \hline 
Estimated using & Model & Null hypothesis & AIC & BIC & KS-test & AIC & BIC & KS-test\\ \hline
\multirow{10}{*}{Gaussian Likelihood} 
& \multirow{2}{*}{$\Lambda$CDM}   & Gaussian  & 1509.14 & 1517.66 & 0.15 & 1469.76 & 1478.28 & 0.89 \\ \cline{3-9}
&        & Logistic  & 1497.02 & 1505.54 & 0.92 & 1470.55 & 1479.07 & 0.98 \\ \cline{2-9}
& \multirow{2}{*}{wCDM}   & Gaussian  & 1510.45 & 1518.97 & 0.16 & 1471.49 & 1480.01 & 0.83 \\ \cline{3-9}
&        & Logistic  & 1498.40 & 1506.92 & 0.86 & 1472.96 & 1481.48 & 0.92 \\ \cline{2-9}
& \multirow{2}{*}{CPL}  & Gaussian  & 1513.56 & 1522.08 & 0.14 & 1477.69 & 1486.22 & 0.72 \\ \cline{3-9}
&        & Logistic  & 1501.80 & 1510.33 & 0.88 & 1479.41 & 1487.94 & 0.91 \\ \cline{2-9}
& \multirow{2}{*}{JBP}   & Gaussian  & 1510.24 & 1518.76 & 0.12 & 1473.14 & 1481.66 & 0.65 \\ \cline{3-9}
&        & Logistic  & 1498.15 & 1506.67 & 0.87 & 1474.49 & 1483.01 & 0.83 \\ \cline{2-9}
& \multirow{2}{*}{Log}   & Gaussian  & 1510.58 & 1519.11 & 0.17 & 1474.30 & 1482.82 & 0.71 \\ \cline{3-9}
&        & Logistic  & 1498.53 & 1507.06 & 0.90 & 1475.73 & 1484.25 & 0.90 \\ \cline{2-9}
& \multirow{2}{*}{Linear} & Gaussian  & 1516.22 & 1524.75 & 0.06 & 1480.94 & 1489.46 & 0.65 \\ \cline{3-9}
&        & Logistic  & 1504.78 & 1513.31 & 0.55 & 1483.00 & 1491.52 & 0.90 \\ \hline

\multirow{10}{*}{Logistic Likelihood}
& \multirow{2}{*}{$\Lambda$CDM}   & Gaussian  & 1504.00 & 1512.52 & 0.11 & 1460.90 & 1469.42 & 0.86 \\ \cline{3-9}
&        & Logistic  & 1490.48 & 1499.00 & 0.78 & 1459.80 & 1468.32 & 0.82 \\ \cline{2-9}
& \multirow{2}{*}{wCDM}   & Gaussian  & 1502.49 & 1511.01 & 0.17 & 1457.75 & 1466.27 & 0.91 \\ \cline{3-9}
&        & Logistic  & 1488.63 & 1497.15 & 0.72 & 1457.25 & 1465.77 & 0.98 \\ \cline{2-9}
& \multirow{2}{*}{CPL}  & Gaussian  & 1498.04 & 1506.56 & 0.13 & 1462.49 & 1471.02 & 0.71 \\ \cline{3-9}
&        & Logistic  & 1483.43 & 1491.95 & 0.71 & 1462.57 & 1471.09 & 0.88 \\ \cline{2-9}
& \multirow{2}{*}{JBP}   & Gaussian  & 1499.98 & 1508.51 & 0.12 & 1460.29 & 1468.82 & 0.71 \\ \cline{3-9}
&        & Logistic  & 1485.67 & 1494.19 & 0.60 & 1459.79 & 1468.31 & 0.84 \\ \cline{2-9}
& \multirow{2}{*}{Log}   & Gaussian  & 1502.66 & 1511.19 & 0.18 & 1461.05 & 1469.58 & 0.71 \\ \cline{3-9}
&        & Logistic  & 1488.75 & 1497.27 & 0.74 & 1460.56 & 1469.08 & 0.85 \\ \cline{2-9}
& \multirow{2}{*}{Linear} & Gaussian  & 1494.40 & 1502.92 & 0.11 & 1461.26 & 1469.79 & 0.68 \\ \cline{3-9}
&        & Logistic  & 1479.25 & 1487.77 & 0.89 & 1461.99 & 1470.51 & 0.88 \\ \hline
\end{tabular}}
\caption{Goodness of fit test values for the two equal bins}
\label{goft2b}
\end{table}

\begin{table}[H]
\centering
\scalebox{0.7}{
\begin{tabular}{|l|l|l|l|l|l|l|l|l|l|l|l|}
\hline
    \multicolumn{12}{|c|}{Three Equal Bins} \\ \hline
  \multicolumn{3}{|c|}{}  & \multicolumn{3}{|c|}{First Bin} & \multicolumn{3}{|c|}{Second Bin} & \multicolumn{3}{|c|}{Third Bin} \\ \hline 
Estimated using & Model & Null hypothesis & AIC & BIC & KS-test & AIC & BIC & KS-test & AIC & BIC & KS-test \\ \hline
\multirow{10}{*}{Gaussian Likelihood} 
& \multirow{2}{*}{$\Lambda$CDM}   & Gaussian  & 1002.27 & 1009.98 & 0.42 & 996.36  & 1004.07 & 0.83 & 982.80 & 990.51 & 0.52 \\ \cline{3-12}
&        & Logistic  & 995.26  & 1002.97 & 0.97 & 996.07  & 1003.78 & 0.98 & 978.67 & 986.39 & 0.97 \\ \cline{2-12}
& \multirow{2}{*}{wCDM}   & Gaussian  & 1002.72 & 1010.43 & 0.40 & 1000.67 & 1008.38 & 0.71 & 985.37 & 993.08 & 0.56 \\ \cline{3-12}
&        & Logistic  & 995.82  & 1003.53 & 0.96 & 1000.64 & 1008.35 & 0.98 & 981.02 & 988.74 & 0.88 \\ \cline{2-12}
& \multirow{2}{*}{CPL}  & Gaussian  & 1007.02 & 1014.73 & 0.37 & 991.29  & 999.00  & 0.74 & 981.51 & 989.23 & 0.58 \\ \cline{3-12}
&        & Logistic  & 1000.70 & 1008.41 & 0.93 & 990.63  & 998.34  & 0.99 & 976.69 & 984.40 & 0.93 \\ \cline{2-12}
& \multirow{2}{*}{JBP}   & Gaussian  & 1002.51 & 1010.22 & 0.38 & 1001.62 & 1009.33 & 0.68 & 986.83 & 994.55 & 0.56 \\ \cline{3-12}
&        & Logistic  & 995.57  & 1003.28 & 0.96 & 1001.60 & 1009.31 & 0.98 & 982.31 & 990.03 & 0.80 \\ \cline{2-12}
& \multirow{2}{*}{Log}   & Gaussian  & 1002.70 & 1010.41 & 0.40 & 1000.34 & 1008.05 & 0.73 & 985.40 & 993.11 & 0.42 \\ \cline{3-12}
&        & Logistic  & 995.80  & 1003.51 & 0.96 & 1000.27 & 1007.98 & 0.98 & 980.77 & 988.49 & 0.83 \\ \cline{2-12}
& \multirow{2}{*}{Linear} & Gaussian  & 1064.72 & 1072.43 & 0.73 & 1056.79 & 1064.50 & 0.93 & 1049.97 & 1057.68 & 0.96 \\ \cline{3-12}
&        & Logistic  & 1063.53 & 1071.24 & 1.00 & 1059.35 & 1067.06 & 1.00 & 1052.43 & 1060.15 & 0.94 \\ \hline

\multirow{10}{*}{Logistic Likelihood}
& \multirow{2}{*}{$\Lambda$CDM}   & Gaussian  & 1000.00 & 1007.71 & 0.36 & 974.95  & 982.66  & 0.65 & 980.09 & 987.81 & 0.67 \\ \cline{3-12}
&        & Logistic  & 991.98  & 999.69  & 0.98 & 971.76  & 979.47  & 0.99 & 975.61 & 983.32 & 0.99 \\ \cline{2-12}
& \multirow{2}{*}{wCDM}   & Gaussian  & 999.48  & 1007.19 & 0.28 & 975.02  & 982.73  & 0.72 & 980.54 & 988.26 & 0.68 \\ \cline{3-12}
&        & Logistic  & 991.37  & 999.08  & 0.92 & 971.97  & 979.68  & 1.00 & 975.69 & 983.41 & 0.95 \\ \cline{2-12}
& \multirow{2}{*}{CPL}  & Gaussian  & 999.35  & 1007.06 & 0.37 & 976.64  & 984.35  & 0.56 & 983.87 & 991.58 & 0.56 \\ \cline{3-12}
&        & Logistic  & 990.96  & 998.67  & 0.86 & 973.99  & 981.70  & 0.98 & 978.56 & 986.28 & 0.77 \\ \cline{2-12}
& \multirow{2}{*}{JBP}   & Gaussian  & 998.81  & 1006.52 & 0.23 & 975.06  & 982.77  & 0.70 & 981.68 & 989.39 & 0.48 \\ \cline{3-12}
&        & Logistic  & 990.50  & 998.21  & 0.92 & 971.96  & 979.67  & 1.00 & 976.46 & 984.17 & 0.91 \\ \cline{2-12}
& \multirow{2}{*}{Log}   & Gaussian  & 999.38  & 1007.09 & 0.30 & 975.00  & 982.71  & 0.72 & 981.63 & 989.34 & 0.48 \\ \cline{3-12}
&        & Logistic  & 991.24  & 998.95  & 0.93 & 971.93  & 979.64  & 1.00 & 976.43 & 984.15 & 0.89 \\ \cline{2-12}
& \multirow{2}{*}{Linear} & Gaussian  & 998.22  & 1005.93 & 0.35 & 977.99  & 985.70  & 0.56 & 985.46 & 993.18 & 0.43 \\ \cline{3-12}
&        & Logistic  & 989.42  & 997.13  & 0.88 & 975.29  & 983.00  & 0.92 & 980.32 & 988.04 & 0.83 \\ \hline
\end{tabular}}
\caption{Goodness of fit test values for the three equal bins}
\label{goft3b}
\end{table}

\begin{table}[H]
\centering
\scalebox{0.8}{
\begin{tabular}{|l|l|l|l|l|l|l|l|l|l|l|l|}
\hline
    \multicolumn{12}{|c|}{Survey} \\ \hline
  \multicolumn{3}{|c|}{}  & \multicolumn{3}{|c|}{SNLS} & \multicolumn{3}{|c|}{SDSS} & \multicolumn{3}{|c|}{PS1} \\ \hline 
Estimated using & Model & Null hypothesis & AIC & BIC & KS-test & AIC & BIC & KS-test & AIC & BIC & KS-test \\ \hline
\multirow{10}{*}{Gaussian Likelihood} 
& \multirow{2}{*}{$\Lambda$CDM}   & Gaussian  & 663.52 & 670.45 & 0.47 & 968.48  & 976.11 & 0.39 & 774.26 & 781.52 & 0.68 \\ \cline{3-12}
&        & Logistic  & 659.39 & 666.32 & 0.91 & 963.82  & 971.45 & 0.91 & 772.21 & 779.47 & 0.95 \\ \cline{2-12}
& \multirow{2}{*}{wCDM}   & Gaussian  & 665.21 & 672.14 & 0.58 & 971.76  & 979.39 & 0.32 & 777.69	& 784.95 & 0.76 \\ \cline{3-12}
&        & Logistic  & 661.49 & 668.42 & 0.98 & 966.86  & 974.49 & 0.93 & 776.11 & 783.37 & 0.99\\ \cline{2-12}
& \multirow{2}{*}{CPL}  & Gaussian  & 667.11 & 674.03 & 0.66 & 976.34  & 983.97 & 0.43 & 775.24 & 782.50 & 0.43\\ \cline{3-12}
&        & Logistic  & 663.68 & 670.61 & 0.99 & 971.32  & 978.95 & 0.95 & 773.38 & 780.64 & 0.95 \\ \cline{2-12}
& \multirow{2}{*}{JBP}   & Gaussian  & 666.04 & 672.96 & 0.50 & 971.76  & 979.38 & 0.33 & 780.91 & 788.18 & 0.74\\ \cline{3-12}
&        & Logistic  & 662.18 & 669.11 & 0.96 & 966.71  & 974.34 & 0.97 & 779.83 & 787.09 & 0.99 \\ \cline{2-12}
& \multirow{2}{*}{Log}   & Gaussian  & 665.66 & 672.59 & 0.49 & 971.71  & 979.33 & 0.30 & 777.07 & 784.33 & 0.69\\ \cline{3-12}
&        & Logistic  & 661.81 & 668.74 & 0.97 & 966.78  & 974.40 & 0.94 & 775.46 & 782.73 & 0.98\\ \cline{2-12}
& \multirow{2}{*}{Linear} & Gaussian  & 668.18 & 675.10 & 0.84 & 980.34  & 987.97 & 0.44 & 843.20 & 850.47 & 0.93\\ \cline{3-12}
&        & Logistic  & 665.35 & 672.28 & 0.98 & 975.11  & 982.74 & 0.99 & 846.87 & 854.13 & 0.72 \\ \hline

\multirow{10}{*}{Logistic Likelihood}
& \multirow{2}{*}{$\Lambda$CDM}   & Gaussian  & 659.38 & 666.31 & 0.71 & 936.61  & 944.24 & 0.29 & 768.18 & 775.44 & 0.79 \\ \cline{3-12}
&        & Logistic  & 653.94 & 660.87 & 0.98 & 922.17  & 929.80 & 0.96 & 766.21 & 773.47 & 0.99\\ \cline{2-12}
& \multirow{2}{*}{wCDM}   & Gaussian  & 658.90 & 665.83 & 0.68 & 936.44  & 944.07 & 0.36 & 769.09 & 776.35 & 0.96\\ \cline{3-12}
&        & Logistic  & 653.55 & 660.48 & 0.98 & 922.32  & 929.95 &  0.91 & 767.43 & 774.69 & 1.00 \\ \cline{2-12}
& \multirow{2}{*}{CPL}  & Gaussian  & 659.49 & 666.42 & 0.68 & 936.48  & 944.11 & 0.28 & 771.59 & 778.85 & 0.39\\ \cline{3-12}
&        & Logistic  & 654.48 & 661.41 & 1.00 & 922.67  & 930.29 & 0.93 & 770.13 & 777.39 & 0.89\\ \cline{2-12}
& \multirow{2}{*}{JBP}   & Gaussian  & 660.06 & 666.99 & 0.59 & 936.70  & 944.33 & 0.27 & 770.11 & 777.37 & 0.89\\ \cline{3-12}
&        & Logistic  & 654.57 & 661.50 & 1.00 & 922.53  & 930.16 & 0.92 & 768.69 & 775.95 & 1.00\\ \cline{2-12}
& \multirow{2}{*}{Log}   & Gaussian  & 659.35 & 666.28 & 0.63 & 936.53  & 944.16 & 0.28 & 768.81 & 776.07 & 0.94 \\ \cline{3-12}
&        & Logistic  & 653.89 & 660.81 & 1.00 & 922.56  & 930.19 & 0.91 & 767.13 & 774.39 & 1.00\\ \cline{2-12}
& \multirow{2}{*}{Linear} & Gaussian  & 658.94 & 665.87 & 0.65 & 936.60  & 944.23 & 0.26 & 806.77 & 814.03 & 0.99\\ \cline{3-12}
&        & Logistic  & 654.70 & 661.63 & 1.00 & 923.19  & 930.82 & 0.85 & 808.49 & 815.75 & 0.96 \\ \hline
\end{tabular}}
\caption{Goodness of fit test values for SNLS and SDSS data}
\label{goftS}
\end{table}

\section{Gelman-Rubin convergence} \label{GRCT}
The convergence of MCMC chains (for each parameter of the model) is given by Gelman-Rubin (GR) convergence ratio. If we have $M$ chains with $N$ steps,  the $m_{th}$ chain can be denoted by $\theta_{1}^m, \theta_{2}^m,..... \theta_{N}^m$. 
For each parameter $\theta$, the intra-chain mean is defined as
\begin{equation}
 \hat{\theta}_m = \frac{1}{N} \sum_{i=1}^N \theta_{i}^m.   
\end{equation}
Hence, the intra-variance $ \sigma_m^2$, inter-chain mean $\hat{\theta} $, chain-to-chain variance $B$, averaged variances of the chain $W$ are  given as:
\begin{align}
     \sigma_m^2 &= \frac{1}{N-1} \sum_{i=1}^N (\theta_{i}^m - \hat{\theta}_m)^2  \\
      \hat{\theta} & = \frac{1}{M} \sum_{m=1}^M \hat{\theta}_m   \\
      B &= \frac{N}{M-1} \sum_{m=1}^M (\hat{\theta}_m-\hat{\theta})^2 \\
      W &= \frac{1}{M} \sum_{m=1}^M \sigma_m^2 . 
\end{align}

We define the unbiased estimator of true variance under convergence $\hat{V}$ as,
\begin{equation}
 \hat{V} = \frac{N-1}{N} W + \frac{M+1}{MN} B   
\end{equation}
If the chains have converged then W is also the unbiased estimator of true variance. Then we should have,
\begin{equation}
 R = \frac{\hat{V}}{W} \approx 1   
\end{equation}
The convergence test results of the Pantheon dataset for the Gaussian and Logistic likelihood function are as follows:
\begin{table}[h]
\centering
\scalebox{0.85}{
    \begin{tabular}{|l|l|l|l|l|l||l|l|l|l|l|}
    \hline  
   &  \multicolumn{5}{|c|}{Gaussian} & \multicolumn{5}{|c|}{Logistic}
   \\ \cline{2-11} 
    Models  &   $R_H$  &    $R_{\Omega_m} $  & $R_{w0}$    & $R_{w^\prime}$     &$R_{w_1}$ &   $R_H$  &    $R_{\Omega_m} $  & $R_{w0}$    & $R_{w^\prime}$     &$R_{w_1}$ \\
    \cline{1-11} 
  $\Lambda CDM $ &0.999    & 0.999 &  
 &       &  & 0.999 & 1.000 & ~ & ~ & ~ \\   \cline{2-11} 
   $w CDM $ &  0.999    &  0.999 & 0.999&    
  & &    1.000 & 1.000 & 1.000 & ~ & ~  \\  
  \cline{2-11} 
  $CPL$ &   0.999   & 0.999 & 0.999 & 0.999  &   &   0.999 & 0.999 & 1.000 & 0.999 & ~ 
\\  \cline{2-11} 
 $JBP$ &    0.999    &  0.999 &  0.999 & 0.999 &  &  0.999 & 0.999 & 0.999 & 0.999 & ~  \\ 
 \cline{2-11} 
  $log$ &    0.999    &  0.999 &  0.999 & 0.999 &  &    0.999 & 0.999 & 0.999 & 0.999 & ~   \\ 
 \cline{2-11} 
 $linear$ &   0.999  & 0.999 & 0.999 & & 0.999 &   0.999 & 0.999 & 0.999 & ~ & 0.999  \\  \hline

		\end{tabular}
  }
		\caption{Gelmin Rubin convergence test for Pantheon dataset}
		\label{GR for Logistic}
\end{table}

The GR convergence ratio for the binned datasets, including the SDSS and SNLS surveys, falls within the range of \(0.999 < R < 1.000\) for all parameterisations and both Gaussian and Logistic likelihood functions. This indicates that the MCMC chains have converged.
%A same pattern is observed for the simulated dataset.

\medskip
\bibliographystyle{JHEP}
\bibliography{jcappub}

\end{document}